\documentclass[twocolumn]{aa}
\usepackage[varg]{txfonts}

\usepackage{graphicx}
\usepackage{natbib}
\usepackage{color}
\usepackage{subfig}
\usepackage[switch]{lineno} 
\usepackage{xspace}
\usepackage{microtype}
\usepackage{amssymb}
\usepackage{mathtools}
\usepackage{multirow}
\def\link_col{blue}
\usepackage{xspace}
\usepackage{url}
\usepackage{wasysym}
\usepackage{caption}

\def\chandra{{\it Chandra}\xspace}

\def\xray{$\rm X$-ray\xspace}
\def\xrays{$\rm X$-rays\xspace}
\def\gray{$\gamma$-ray\xspace}

\def \msun{\mbox{M$_{\odot}$}\xspace}
\def \nh{$N_{\rm H}$\xspace}

\begin{document}

\title{ Energy distribution of relativistic electrons in the  kiloparsec scale jet of M87 with \chandra}
\titlerunning{The Jet of M87}
\author{Xiao-Na Sun\inst{1}
\and Rui-Zhi Yang\inst{1}
\and Frank M. Rieger\inst{1,2}
\and Ruo-Yu Liu\inst{1}
\and Felix Aharonian\inst{1,3,4,5}}
\institute{Max-Planck-Institut f{\"u}r Kernphysik, P.O. Box 103980, 69029 Heidelberg, Germany\\
\email{sun@mpi-hd.mpg.de}\label{inst1}
\and ZAH, Institut f\"ur Theoretische Astrophysik, Heidelberg University, Philosophenweg 12, 69120 Heidelberg, Germany
\and Dublin Institute for Advanced Studies, 31 Fitzwilliam Place, Dublin 2, Ireland
\and Gran Sasso Science Institute, 7 viale Francesco Crispi, 67100 L'Aquila (AQ), Italy
\and MEPHI, Kashirskoe shosse 31, 115409 Moscow, Russia
}

\abstract {The \xray emission from the jets in active galactic nuclei (AGN) carries important information on the distributions of 
relativistic electrons and magnetic fields on large scales. We reanalysed 
archival \chandra observations on the jet of M87 from 2000 
to 2016 with a total exposure of 1460 kiloseconds to explore the \xray emission characteristics along the jet. We investigated the variability 
behaviours of the nucleus and the inner jet component HST-1, and confirm indications for day-scale \xray variability in the nucleus 
contemporaneous to the 2010 high TeV \gray state. HST-1 shows a general decline in \xray flux over the last few years 
consistent with its synchrotron interpretation.
We extracted the \xray spectra for the nucleus and all knots in the jet, showing that they are compatible with a single power law within 
the \xray band. There are indications that the resultant \xray photon index  exhibit a trend, with slight but significant index variations 
ranging from $\simeq 2.2$ (e.g. in knot D) to $\simeq 2.4-2.6$ (in the outer knots F, A, and B). When viewed in a multiwavelength 
context, a more complex situation  arises. 
Fitting the radio to \xray spectral energy distributions (SEDs) assuming a synchrotron 
origin, we show that a broken power-law electron spectrum with break energy $E_b$ around $1~(300\mu G/B)^{1/2}$ TeV allows 
a satisfactory description of the multiband SEDs for most of the knots. However, in the case of knots B, C, and D we find indications 
that an additional high-energy component is needed to adequately reproduce the broad-band SEDs. We discuss the implications and 
suggest that a stratified jet model may account for the differences.}
\maketitle

\section{Introduction}
\label{sec:intro}
M87 (Virgo A, NGC 4486, 3C 274), the giant elliptical galaxy located in the Virgo cluster at a distance of $16.7 \pm 0.6~{\rm Mpc}$ 
($1\arcsec=78~\rm pc$), is one of the closest radio galaxies  \citep{Blakeslee09, Jordan05}. It is known to host a central black hole 
with a mass of $\simeq(3-6)\times 10^{9}$\,\msun \citep{Macchetto97,Gebhardt11,Walsh13} and a one-sided $\sim$ 30\arcsec \ scale 
jet \citep{Marshall02, Harris03}.
The jet is characterised by a viewing angle between $10^\circ-25^\circ$ and reveals superluminal motion of optical components of 
$(4-6)$c within 6\arcsec\ of the nucleus based on HST observations \citep{Biretta99}.
Its jet power $P_{\rm j}$ is somewhat uncertain with estimates ranging from a few times $10^{43}~\rm erg\, s^{-1}$ up to $10^{45}~\rm 
erg\, s^{-1}$  \citep[e.g.][]{Reynolds96,deGasperin12,Mo16,Levinson17}. 
Owing to its proximity and high surface brightness at radio wavelengths and above, M87 has become a key laboratory to investigate 
the property of relativistic jets \citep[e.g.][]{Doeleman12,Hada16,Mertens16,Britzen17}.

Both the nucleus and several bright jet knots have been detected at radio, optical, and \xray wavelengths. Their broad-band SEDs have 
been extensively studied \citep[e.g.][]{Biretta91, Sparks96, Perlman01, Zhang10}. The high-resolution 
observations performed by \chandra in the \xray band make it possible to investigate the SEDs of the jet substructures in or close to 
the spectral  cut-off regions, imposing important constraints on radiation models \citep[e.g.][]{Marshall02, Wilson02,Perlman05,Waters05, 
Zhang10}.  

Owing to the accumulative exposure and the recently enhanced software tools of \chandra, an improved analysis can now be performed
to derive more accurate spectrometric information and to further investigate radiation mechanisms in this region.  
In this paper we present the results of a detailed analysis of the \chandra data on the M87 nucleus and knots observed from 2000 to 2016. 
The paper is structured as follows. In Section~\ref{sec:chandra_analy} we describe details of the \chandra data reduction process, light curve 
analysis, and spectral analysis. In Section~\ref{sec:fitting} we construct the radio to \xray SEDs of the substructures (knots) in M87 and 
provide model fits to infer the spectral distributions of the parent particles. We discuss the consequences in Section~\ref{sec:conc}.

\begin{table*}
\caption{\chandra Observations of M87.}
\centering
\begin{tabular}{cccc|cccc}
\hline\hline
ObsID & Frame Time & Exp Time & Start Date & ObsID & Frame Time & Exp Time & Start Date \\
& (s) & (ks) & (YYY-MM-DD) &&  (s) & (ks) & (YYY-MM-DD) \\ [0.1cm]
\hline
1808&0.4&12.85&2000-07-30&8578&0.4&4.71&2008-04-01 \\
3085&0.4&4.89&2002-01-16 &8579&0.4&4.71&2008-05-15 \\
3084&0.4&4.65&2002-02-12 &8580&0.4&4.7&2008-06-24\\
3086&0.4&4.62&2002-03-30 &8581&0.4&4.66&2008-08-07 \\
3087&0.4&4.97&2002-06-08 &10282&0.4&4.7&2008-11-17 \\
3088&0.4&4.71&2002-07-24 &10283&0.4&4.7&2009-01-07 \\
3975&0.4&5.29&2002-11-17&10284&0.4&4.7&2009-02-20\\
3976&0.4&4.79&2002-12-29 &10285&0.4&4.66&2009-04-01 \\
3977&0.4&5.28&2003-02-04 &10286&0.4&4.68&2009-05-13 \\
3978&0.4&4.85&2003-03-09 &10287&0.4&4.7&2009-06-22\\
3979&0.4&4.49&2003-04-14 &10288&0.4&4.68&2009-12-15 \\
3980&0.4&4.79&2003-05-18 &11512&0.4&4.7&2010-04-11 \\
3981&0.4&4.68&2003-07-03 &11513&0.4&4.7&2010-04-13 \\
3982&0.4&4.84&2003-08-08&11514&0.4&4.53&2010-04-15 \\
4917&0.4&5.03&2003-11-11 &11515&0.4&4.7&2010-04-17 \\
4918&0.4&4.68&2003-12-29 &11516&0.4&4.71&2010-04-20 \\
4919&0.4&4.7&2004-02-12 &11517&0.4&4.7&2010-05-05 \\
4921&0.4&5.25&2004-05-13 &11518&0.4&4.4&2010-05-09 \\
4922&0.4&4.54&2004-06-23 &11519&0.4&4.71&2010-05-11 \\
4923&0.4&4.63&2004-08-05 &11520&0.4&4.6&2010-05-14\\
5737&0.4&4.21&2004-11-26 &13964&0.4&4.54&2011-12-04 \\
5738&0.4&4.67&2005-01-24 &13965&0.4&4.6&2012-02-25\\
5739&0.4&5.15&2005-02-14 &14974&0.4&4.6&2012-12-12\\
5740&0.4&4.7&2005-04-22 &14973&0.4&4.4&2013-03-12\\
5744&0.4&4.7&2005-04-28 &16042&0.4&4.62&2013-12-26 \\
5745&0.4&4.7&2005-05-04 &16043&0.4&4.6&2014-04-02\\
5746&0.4&5.14&2005-05-13&17056&0.4&4.6&2014-12-17 \\
5747&0.4&4.7&2005-05-22 &17057&0.4&4.6&2015-03-19 \\
5748&0.4&4.7&2005-05-30 &18233&0.4&37.25&2016-02-23\\
5741&0.4&4.7&2005-06-03 &18781&0.4&39.51&2016-02-24\\
5742&0.4&4.7&2005-06-21 &18782&0.4&34.07&2016-02-26\\
5743&0.4&4.67&2005-08-06 &18809&0.4&4.52&2016-03-12\\
6299&0.4&4.65&2005-11-29 &18810&0.4&4.6&2016-03-13 \\
6300&0.4&4.66&2006-01-05 &18811&0.4&4.6&2016-03-14 \\
6301&0.4&4.34&2006-02-19 &18812&0.4&4.4&2016-03-16 \\
6302&0.4&4.7&2006-03-30 &18813&0.4&4.6&2016-03-17 \\
6303&0.4&4.7&2006-05-21 &18783&0.4&36.11&2016-04-20 \\
6304&0.4&4.68&2006-06-28 &18232&0.4&18.2&2016-04-27 \\
6305&0.4&4.65&2006-08-02 &18836&0.4&38.91&2016-04-28 \\
7348&0.4&4.54&2006-11-13 &18837&0.4&13.67&2016-04-30 \\
7349&0.4&4.68&2007-01-04&18838&0.4&56.29&2016-05-28 \\
7350&0.4&4.66&2007-02-13&18856&0.4&25.46&2016-06-12 \\ \cline{5-8}
8510&0.4&4.7&2007-02-15&517&3.2&6.99&2000-04-15 \\
8511&0.4&4.7&2007-02-18 &241&3.2&38.04&2000-07-17 \\
8512&0.4&4.7&2007-02-21 &352&3.2&37.68&2000-07-29 \\
8513&0.4&4.7&2007-02-24 &3717&3.2&20.56&2002-07-05 \\
8514&0.4&4.47&2007-03-12 &2707&3.2&98.66&2002-07-06 \\
8515&0.4&4.7&2007-03-14 &6186&3.2&51.55&2005-01-31 \\
8516&0.4&4.68&2007-03-19 &7212&3.1&65.25&2005-11-14 \\
8517&0.4&4.67&2007-03-22 &7210&3.1&30.71&2005-11-16 \\
7351&0.4&4.68&2007-03-24 &7211&3.1&16.62&2005-11-16 \\
7352&0.4&4.59&2007-05-15 &5828&3.1&32.99&2005-11-17 \\
7353&0.4&4.54&2007-06-25 &6186&3.1&51.55&2005-01-31\\
7354&0.4&4.71&2007-07-31 &5827&3.1&156.2&2005-05-05 \\
8575&0.4&4.68&2007-11-25&5826&3.1&126.76&2005-03-03\\ 
8576&0.4&4.69&2008-01-04 &&&&\\ 
8577&0.4&4.66&2008-02-16 &&&&\\
\hline
\end{tabular}
\label{table:chandra_obs}
\end{table*}

\section{\chandra data analysis}\label{sec:chandra_analy}
The {\it \chandra \xray Observatory} ({\it CXO}), launched in 1999, provides unprecedented angular resolution $\le 0.5 \arcsec$ \xray imaging 
and spectroscopy over the energy band 0.1--10 keV (15--0.12 nm) \citep{Weisskopf00, Weisskopf02}.
Its Science Instrument Module (SIM) holds the two focal plane instruments, the Advanced CCD Imaging Spectrometer (ACIS) and the High 
Resolution Camera (HRC). The ACIS is used for spectroscopic studies in the energy range 0.2 -- 10 keV. The HRC accurately records the 
position, number, and energy of \xrays, and images over the range of 0.1 -- 10 keV. 
$\footnote{\url{http://chandra.si.edu/about/science_instruments.html}}$
In this paper, the \chandra data reduction and spectra extraction are performed using CIAO (v4.8) tool and the \chandra Calibration Database 
(CALDB, v4.7.2). We perform the spectral analysis with {\it Sherpa}\footnote{\url{http://cxc.harvard.edu/sherpa/threads/index.html}}.

\subsection{Data preparation} 
\begin{figure}
\centering
\includegraphics[angle=0,scale=0.5]{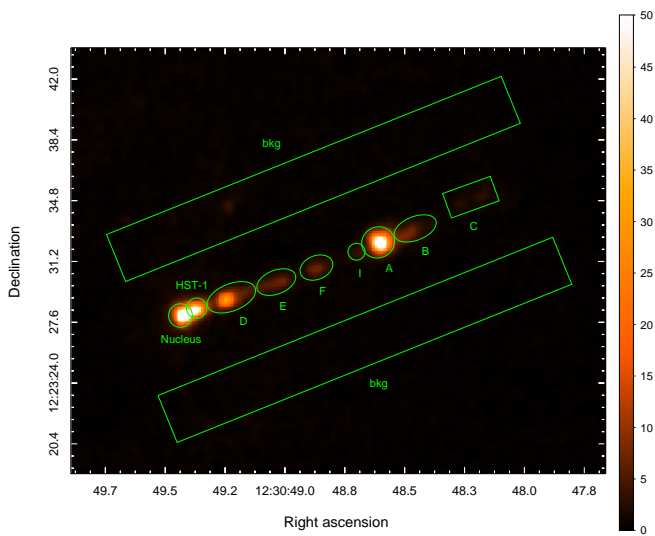}
\caption{Image from the observation on  July 30, 2000 
(ObsID 1808), in the 0.2--10 keV, binned in 0\farcs123 per pixel and smoothed with a Gaussian 
of FWHM=0\farcs3. The source and background regions are indicated by green shapes. Units of right ascension are hours:minutes:seconds, and 
units of declination are degrees:arcminutes:arcseconds.}
\label{fig:countsmap}
\end{figure}

We collected the \chandra ACIS timed exposure (TE) mode observations from April 15, 2000 (ObsID 517), 
to June 12, 2016 (ObsID 18856), to perform a detailed analysis of the M87 jet and nucleus. In Table~\ref{table:chandra_obs} we list all the 
observations, including 99 observations with a 0.4~s frame time and 13 observations with a 3.2/3.1~s frame time. The total exposure time is 
over 1460 kiloseconds.
Figure~\ref{fig:countsmap} provides an exemplary case, showing the ObsID 1808 \xray image binned into 0\farcs123 per pixel and smoothed with 
a Gaussian of FWHM=0.3". 

%
To avoid pile-up \citep[i.e. two or more photons arriving at the same pixel during a single frame time, thus mimicking a single event with the sum of 
the energies; see e.g.][]{Davis01}, we generally selected the observations with frame time of 0.4 s for all substructures analyses. However, because  knots 
E, F, I, B, and C, as shown in Figure~\ref{fig:countsmap}, are much fainter than the nucleus, HST-1, knot D, and knot A, we also included the observations 
with frame time of 3.2/3.1 s in the analysis of these knots.

To reduce uncertainties caused by the position offsets of different observations, we performed the astrometric corrections as follows. In order to obtain 
more reliable source localisation during the process of astrometric correction, the effect of broad-band energy on the exposure map needs to be avoided.  
Thus, we first produced the exposure-corrected image, the weighted exposure map, and the weighted PSF map. 
%
Finally, we performed {\it wcs\_match} and {\it wcs\_update} to match all selected \chandra observations with a reference observation ObsID 1808 separately.
We defined the source region of each knot based on the \xray positions in Table 1 of \citet{Perlman05}. The source 
and background regions are shown in Figure~\ref{fig:countsmap}.


\begin{figure*}
\includegraphics[width=1.\textwidth, height=180px]{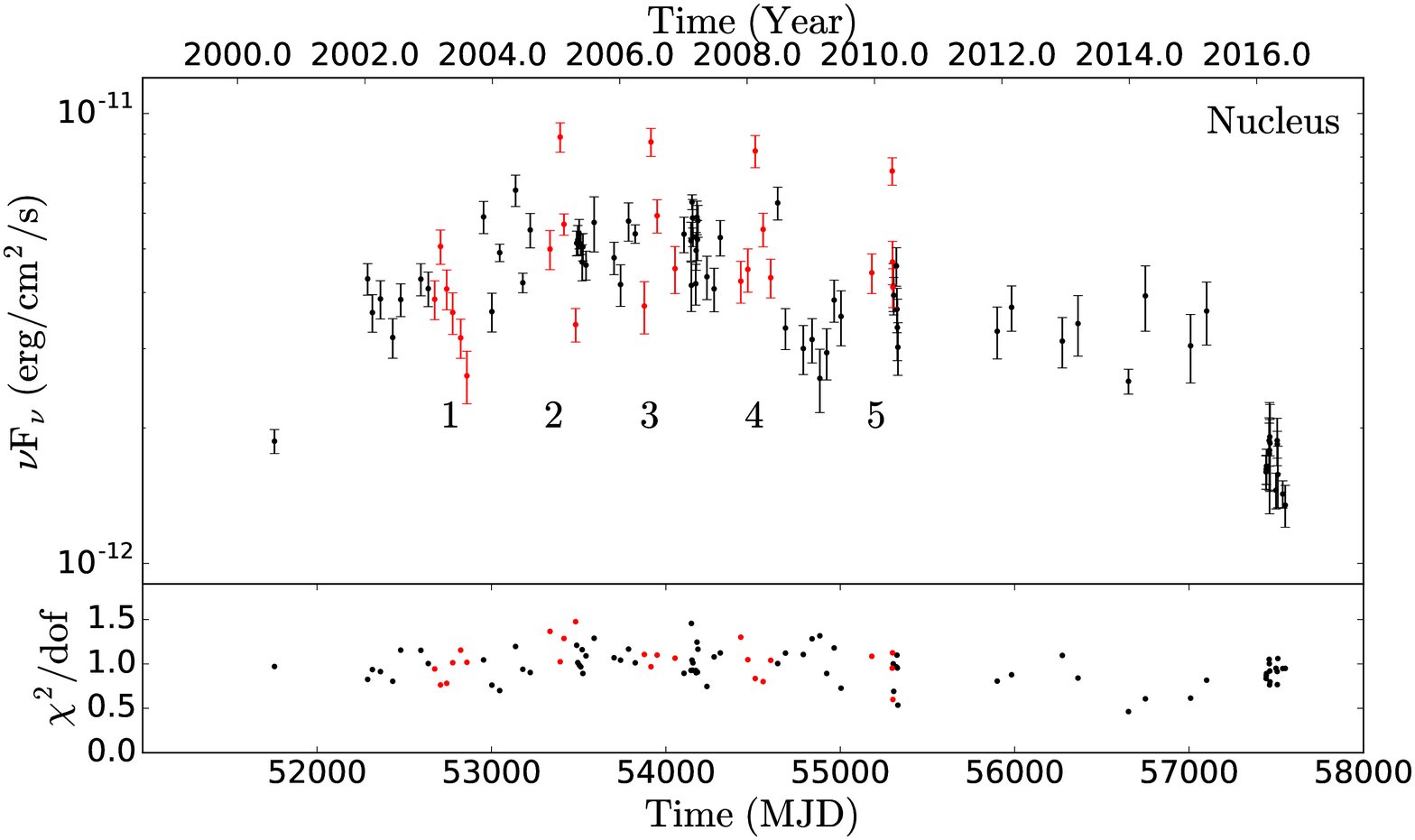}
\includegraphics[width=0.47\textwidth, height=150px]{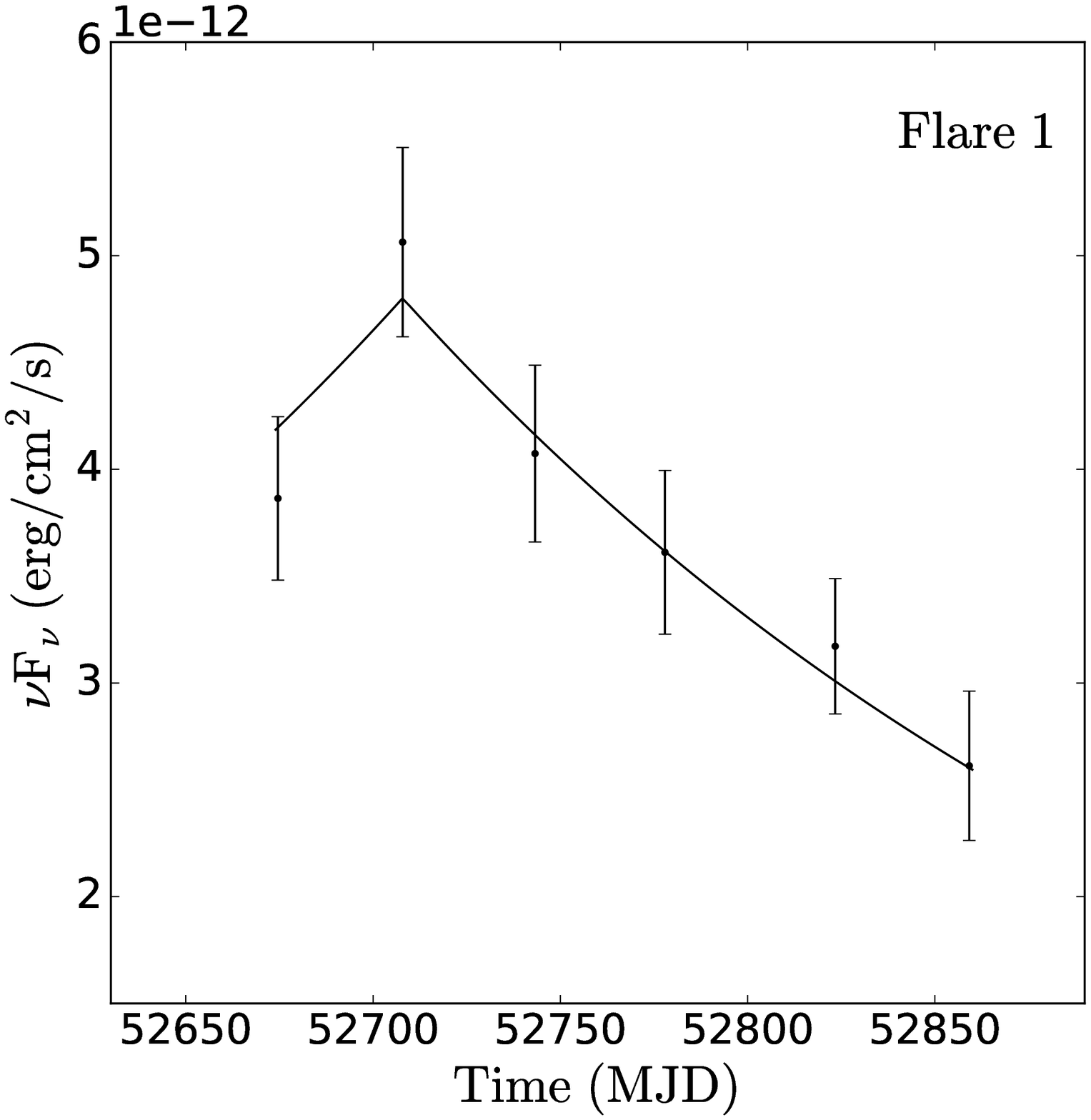}
\includegraphics[width=0.47\textwidth, height=150px]{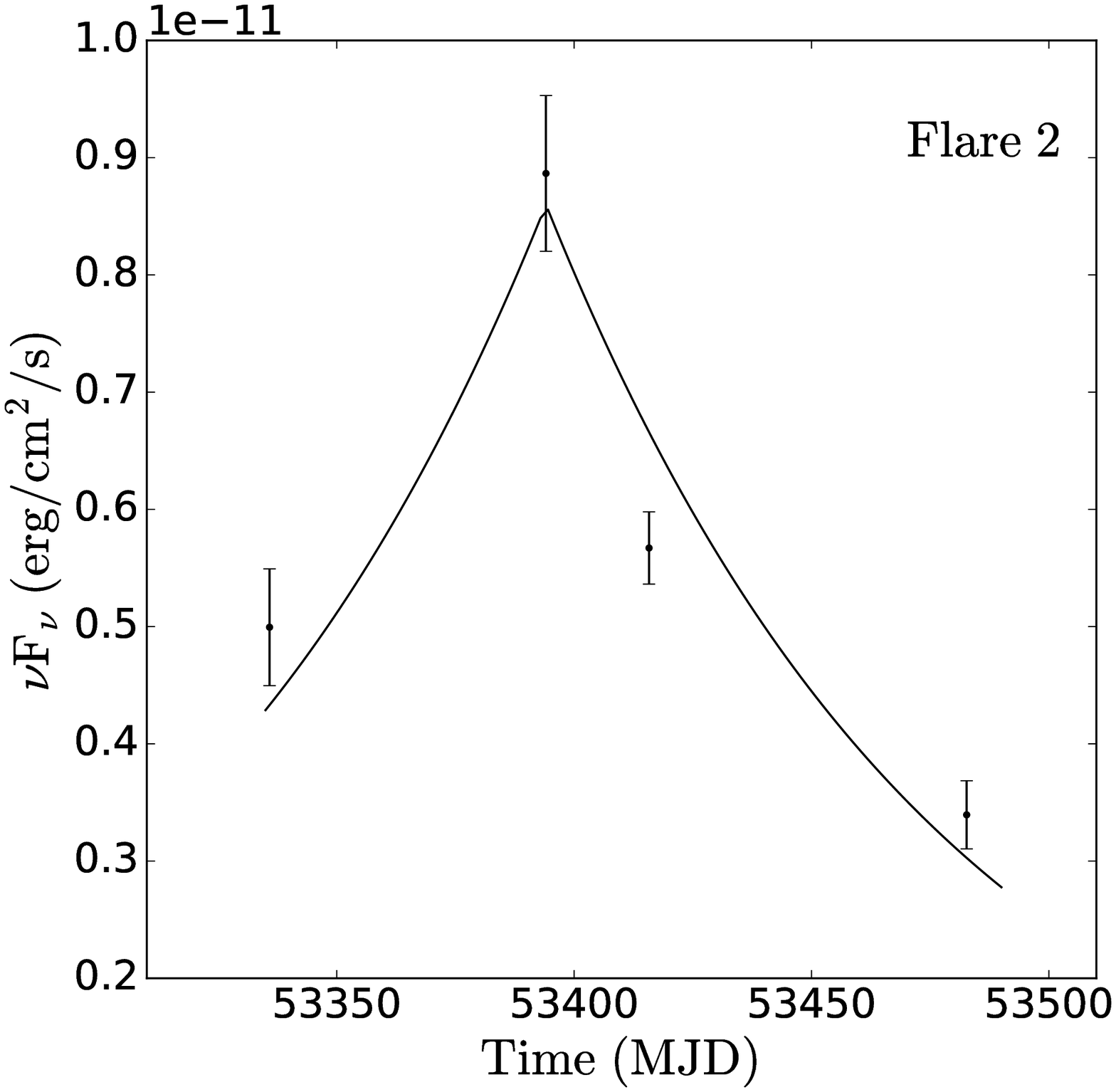}
\includegraphics[width=0.47\textwidth, height=150px]{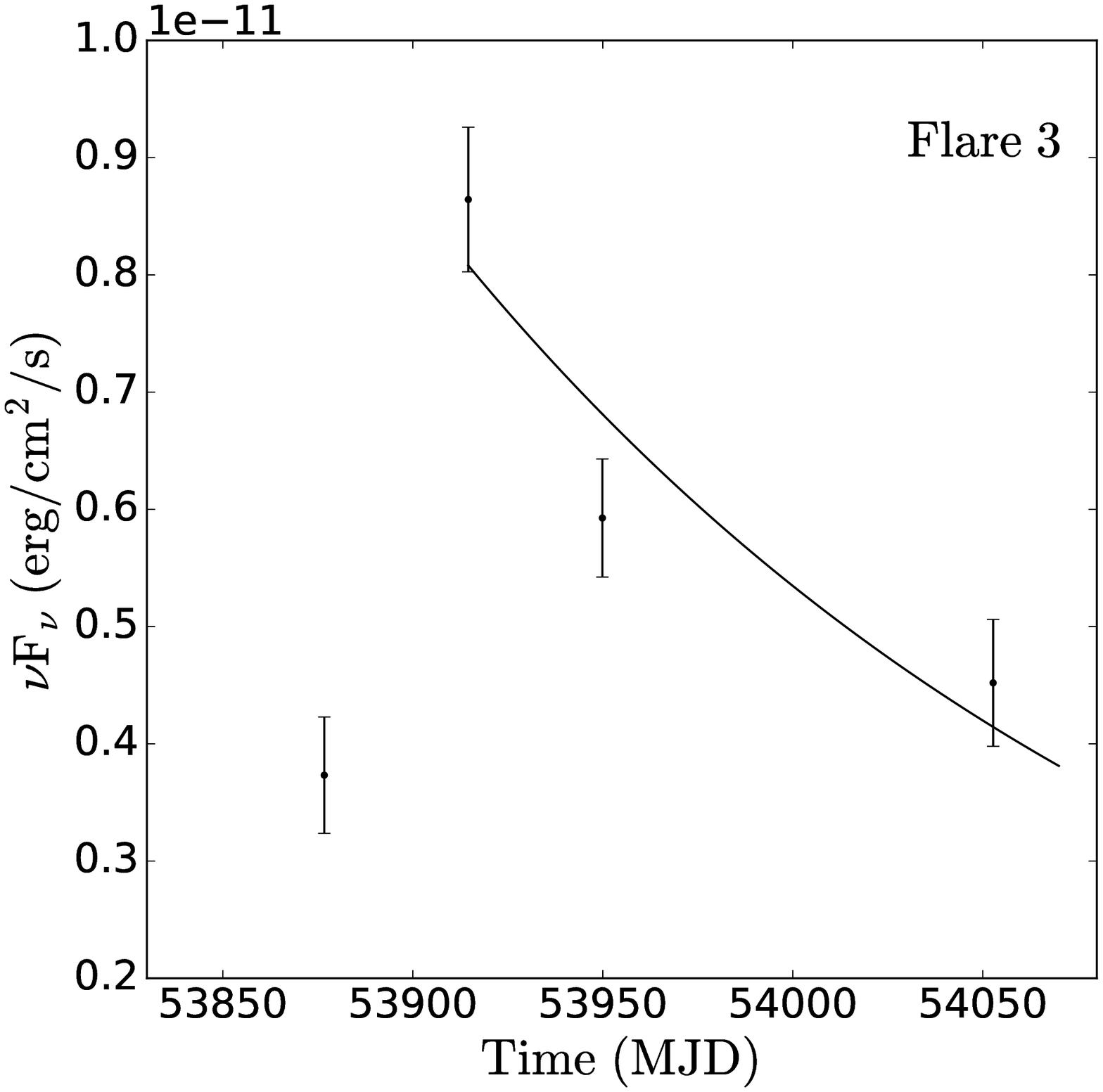}
\includegraphics[width=0.47\textwidth, height=150px]{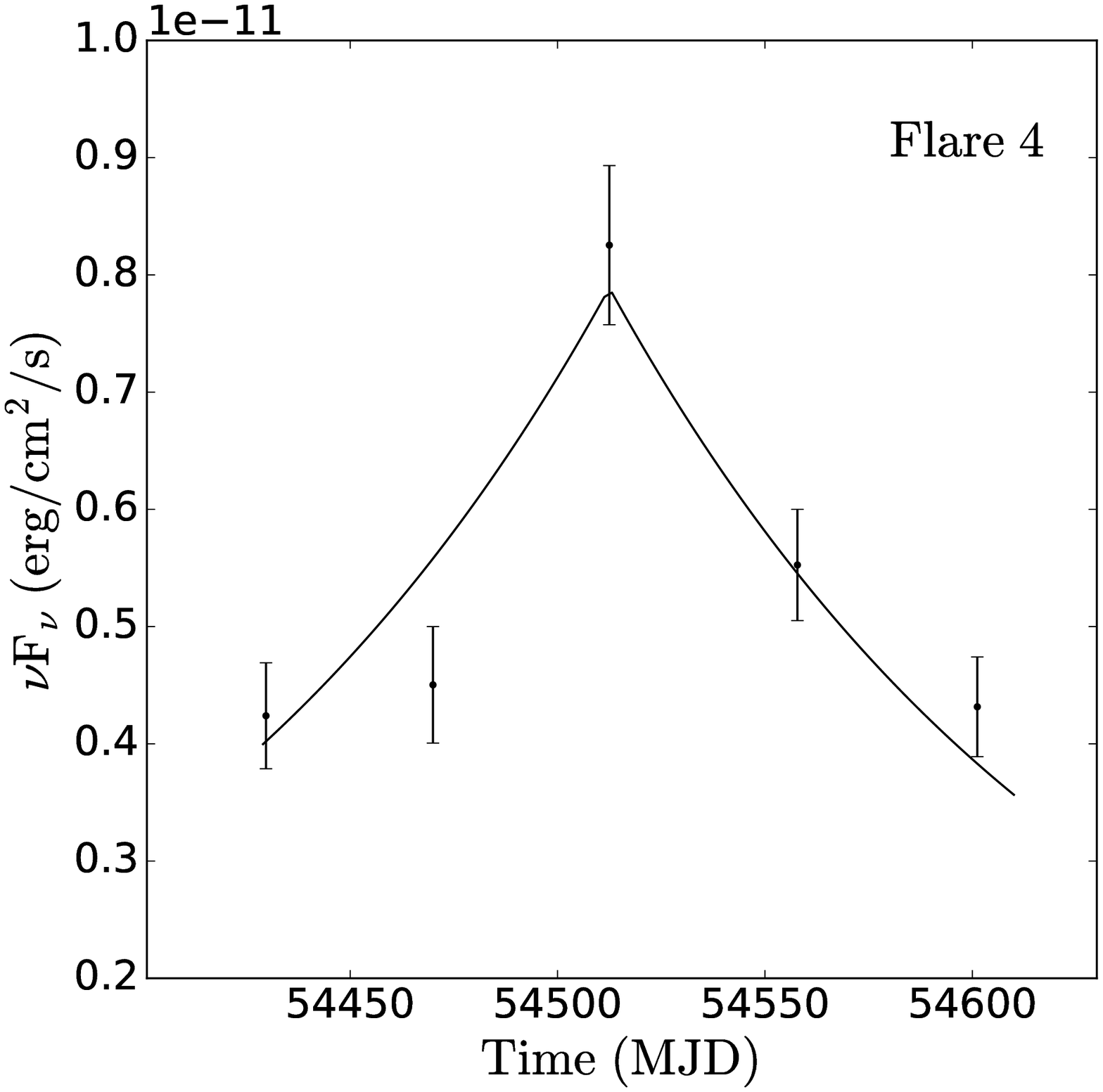}
\includegraphics[width=0.47\textwidth, height=150px]{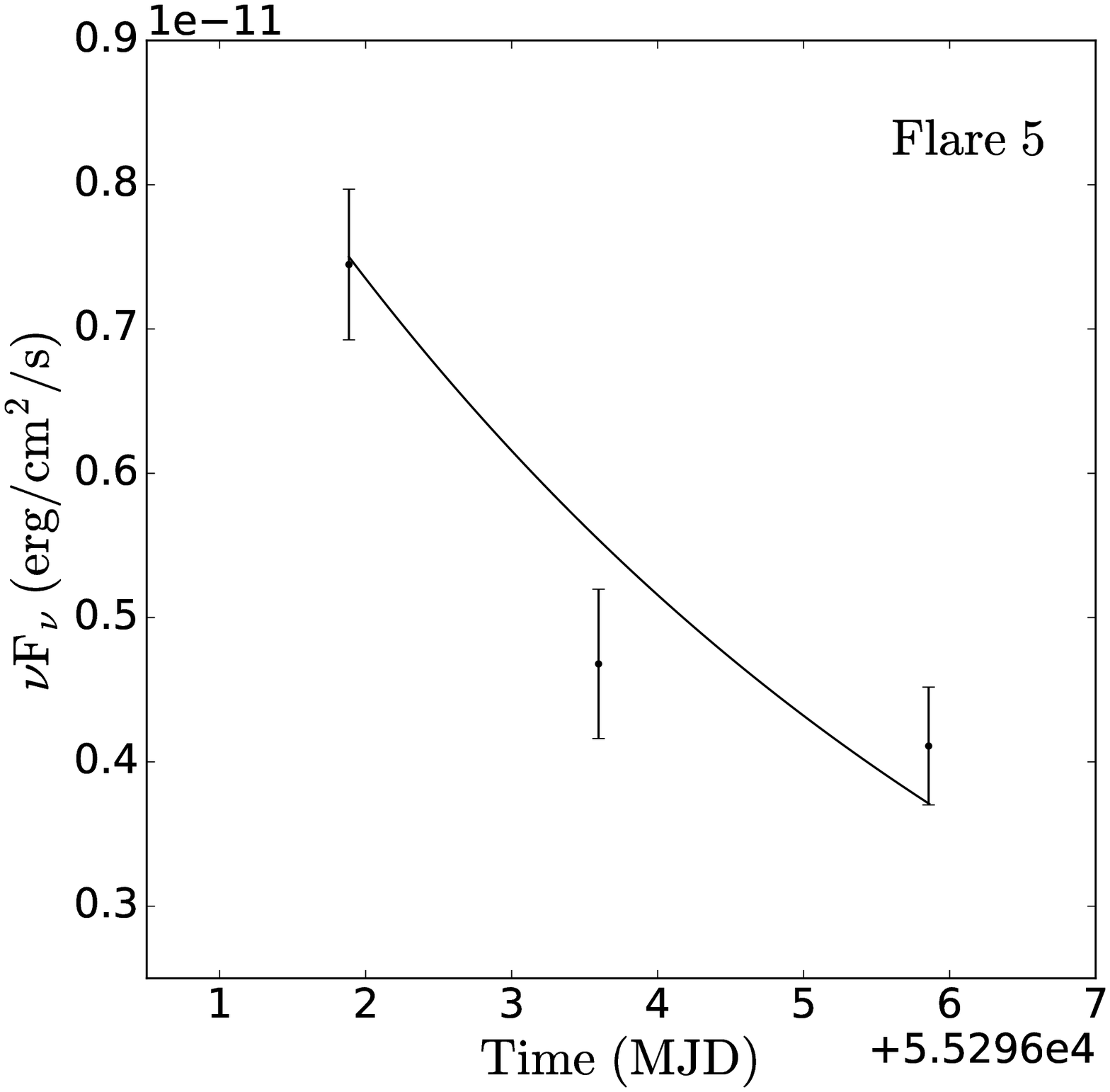}

\caption{Top:  \chandra 0.3--7 keV light curve for the nucleus from  July 29, 2000 (MJD 51754), to  March 17, 2016  (MJD 57464). Five flares (red data 
points, labelled   1--5) are selected for further analysis in the following. Bottom: Zoomed-in light curves of the five flares. The solid lines 
show the fitting results by using Eq.~\ref{equation:flare_fit}.}

\label{fig:lc_nuc}
\end{figure*}

\begin{figure*}
\centering
\includegraphics[width=1.\textwidth, height=180px]{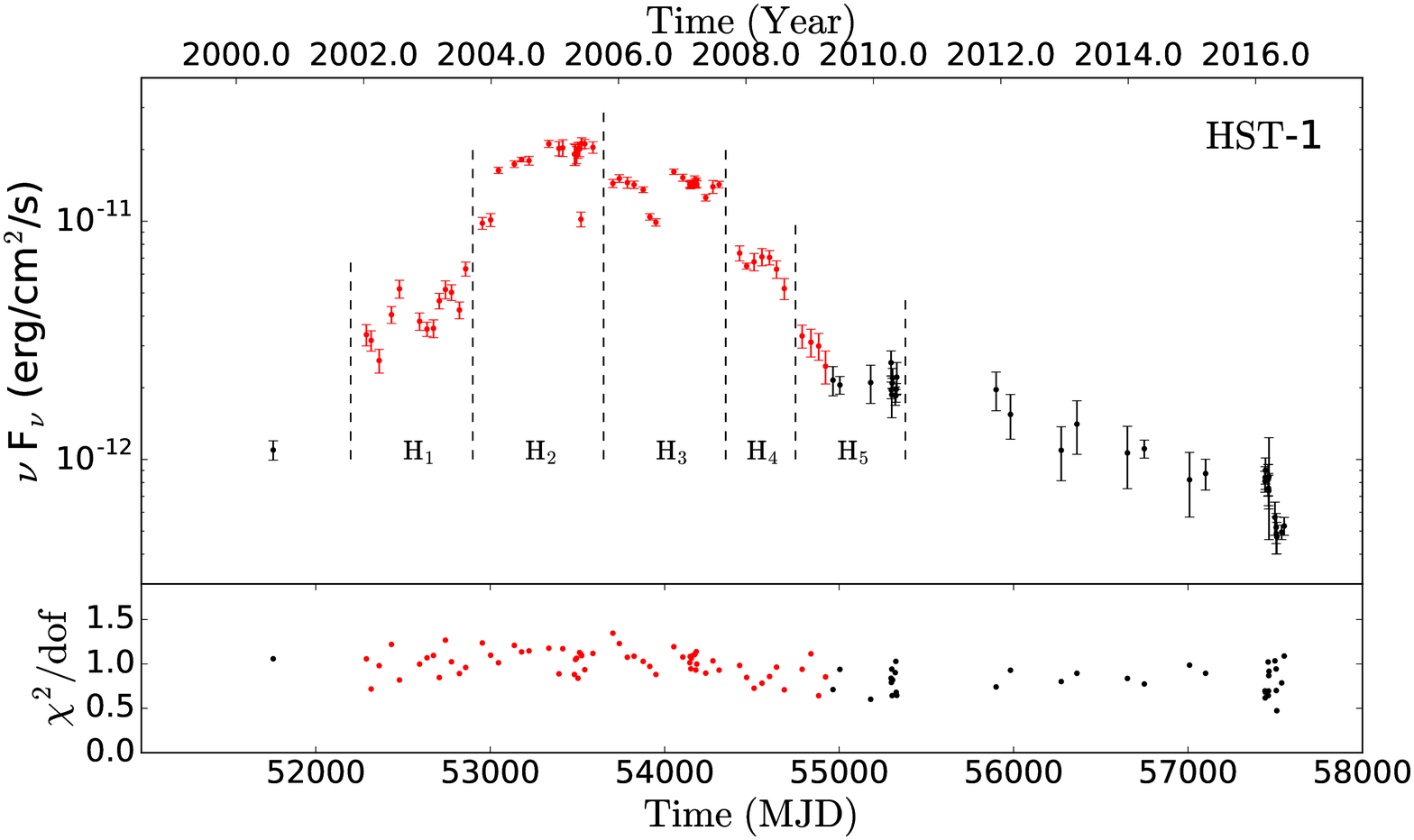}
\caption{ \chandra 0.3--7 keV light curve for HST-1 during the same time interval as in Figure~\ref{fig:lc_nuc}. The black points are considered to belong 
to the low state (L). The dashed lines, on the other hand, separate the high state from January 16, 2002 (MJD 52290), to May 14, 2010 (MJD 55330), into 
five subsections, that are labelled  $\rm H_{1}$, $\rm H_{2}$, $\rm H_{3}$, $\rm H_{4}$, and $\rm H_{5}$. The data from April 01, 2009, to May 14, 2010, 
belong to a transition period between high and low state, and have thus been included in both L and $\rm H_{5}$ calculations.}
\label{fig:lc_hst}
\end{figure*}
\subsection{Time variability}\label{sec:lc_analy}
To study the time variabilities, we extracted the flux for each observation in the 0.3--7 keV energy band and built the light curves for each region.

\subsubsection{Nucleus}
The nucleus reveals significant variabilities as shown in the top panel of Figure ~\ref{fig:lc_nuc}. We defined five flaring periods, labelled 1--5, and present zoomed-in light curves in Figure ~\ref{fig:lc_nuc}.
To investigate the characteristic timescales of these flares in more detail, we fitted the light curves with an exponential function of the form
\begin{equation}
\Phi = \Phi_{0} \times e^{-\vert t-t_{0}\vert/\Delta\tau}\ \rm 
\label{equation:flare_fit}
\end{equation}
Here $\Delta\tau = \tau_{\rm d}/\ln(2)$, $\tau_{\rm d}$ is the characteristic timescales of each flare. We froze $t_{0}$ to the time of the highest data point 
for each flare, and $\Phi_{0}$ and $\tau_{\rm d}$ are free parameters.
In principle, the rising  and decay timescale can be different. The current data set, however, prevents us from deriving the differences of the two timescales due 
to the limited time coverage, especially in the rising stage. Thus, for  flares 1, 2, and 4 we performed fits assuming that both the rising and decaying stage 
have the same $\tau_{\rm d}$. For  flares 3 and 5 we only fit the decaying stage. The corresponding values of $\Phi_{0}$, $\tau_{\rm d}$, and $t_{0}$ are 
listed in Table ~\ref{table:flarederived_pra}. 
The characteristic timescale for flare 5 is shortest with $\tau_d = 3.9\pm1.7$ days. This flare is contemporaneous to the rapid TeV $\gamma$-ray flare seen 
on April 9--10, 2010 (MJD 55296) \citep[see e.g.][]{Abramowski12,harris11}. The day-scale activity seen at \xray and TeV energies in this context favours a 
common physical origin of this emission \citep{Abramowski12,Aliu12}.

\begin{table*}
\caption{ Fitting parameters for the selected flares of the nucleus using a symmetric exponential function.}
\centering
\begin{tabular}{ccccc}
\hline\hline
Flare &Time Range &$\tau_{\rm d}$ & $\rm t_{0}\ (frozen)$ & $\Phi_{0}$ \\

 &(MJD) &(days) &(MJD) & ($\rm \times 10^{-12 }\ \rm erg/cm^2/s$) \\ [0.1cm]
\hline 
Flare 1 &52675-52859&171.27$\pm$23.07&52708&4.80$\pm$0.17 \\[0.1cm]
Flare 2 &53336-53483&66.36$\pm$16.32&53394&8.35$\pm$0.75 \\[0.1cm]
Flare 3 &53877-54053&143.38$\pm$64.76&53915&8.08$\pm$0.97 \\[0.1cm]
Flare 4 &54430-54601&85.11$\pm$18.95&54512&7.89$\pm$0.68 \\[0.1cm]
Flare 5 &55181-55302&3.91$\pm$1.70&55298&7.23$\pm$0.96 \\[0.1cm]
\hline
\end{tabular}
\tablefoot{
All errors are at a  $1\sigma$ confidence level.
}
\label{table:flarederived_pra}
\end{table*}

\subsubsection{HST-1}
HST-1 shows significant variability over time, revealing flux variations much larger than those of the nucleus. During the period from 2002 
to 2010 the total \xray energy flux varied by one order of magnitude, reaching an extreme high state in 2005 (see Fig.~\ref{fig:lc_hst}). 

The 2005 \xray high state of HTS-1 seemingly correlates with a high state of M87 at TeV energies exhibiting rapid day-scale activity 
\citep{Aharonian06}. Early models thus assumed a common physical origin, while the apparent absence of such a correlation for the 
subsequent TeV high state in 2010 has been taken as disfavouring it. We note that there are several reasons why day-scale TeV activity 
related to HST-1 now appears disfavoured \citep[see][for review and discussion]{Rieger12}. The presence or absence of a possible 
\xray-TeV correlation, however, seems rather less conclusive in this regard as changes in the magnetic field and radiating particle 
number in a synchrotron (\xray) and inverse Compton (IC; TeV) approach could fully accommodate both of them.

The \xray flux of HST-1 seems to be continuously decreasing since 2007 with a characteristic decay timescale of $\sim (0.5-1)$ yr. This 
supports previous indications \citep{Harris09}, and while dominant IC cooling is not excluded \citep[e.g.][]{Perlman11}, seems 
compatible with the synchrotron cooling timescale of electrons producing $\sim1$ keV photons in a $\sim$milligauss magnetic field, 
$t_{\rm syn} \lesssim 1.5~(2\times 10^7/\gamma')(1~\rm{mG}/B')^2 \delta^{-1}$ yr. This could be taken as providing some additional 
evidence for a synchrotron origin of the \xray emission.

For the analysis of the spectral variations of HST-1 presented in Section~\ref{sec:spe_analy}, we divide the high-activity period from January 16, 2002 
(MJD 52290), until May 14, 2010 (MJD 55330), into five sections denoted by $\rm H_{1}$, $\rm H_{2}$, $\rm H_{3}$, $\rm H_{4}$, and $\rm H_{5}$.
The observation on July 30, 2000 (ObsID 1808), and the observations after May 13, 2009 (ObsID 10286), are treated as low state (L). The observations 
from April 01, 2009, to May 14, 2010, on the other hand, belong to a transition period between the high and the low state, and have thus been included 
in the following in both, L and $\rm H_{5}$ spectral calculations.
For a bright source, pile-up could lead to distortions in the energy spectrum \citep[e.g.][]{Harris06}. Using the {\it pileup\_map} tool, we estimated the 
pile-up fraction and found that for H2, H3, and H4 the fraction is larger than 10\%, even though we have only selected observations with a frame time of 
0.4 s to avoid  pile-up. Thus, for the following spectral analysis we only include H1, H5, and L for HST-1.

\subsubsection{Other knots}
Knot D and knot A show no significant variability. The flux variations lie within the statistical errors of the flux determination, which is 15\% for knot D and 
5\% for knot A. 
The fluxes derived for  knots E, F, I, B, and C have significant uncertainties due to the limited statistics of the observations. We do not find any 
evidence of variability even after re-binning the observations. With the exception of HST-1, we thus combined all observations for the following spectral 
analysis of the knots. 

\begin{figure*}
\includegraphics[angle=0,scale=0.25]{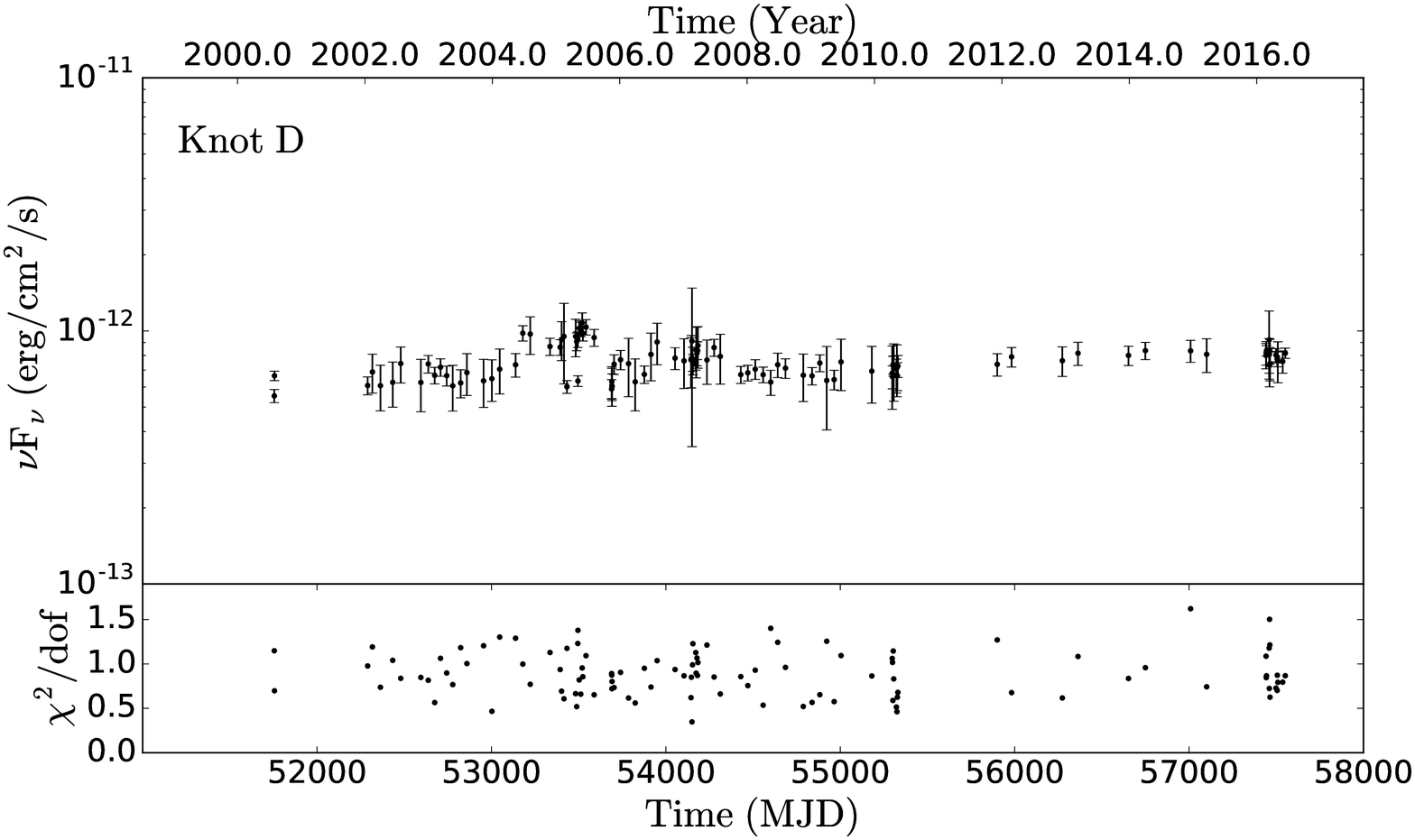}
\includegraphics[angle=0,scale=0.25]{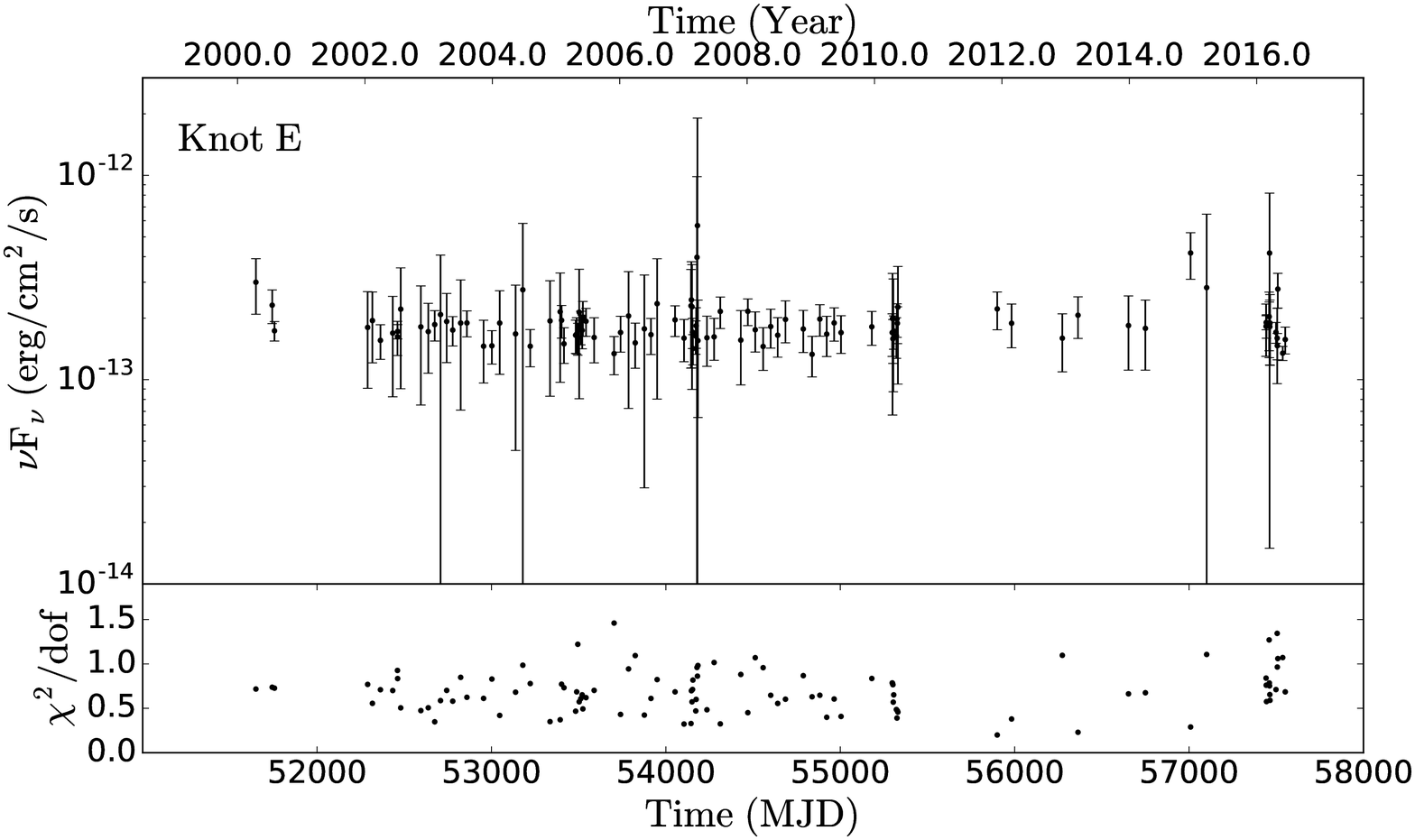}
\includegraphics[angle=0,scale=0.25]{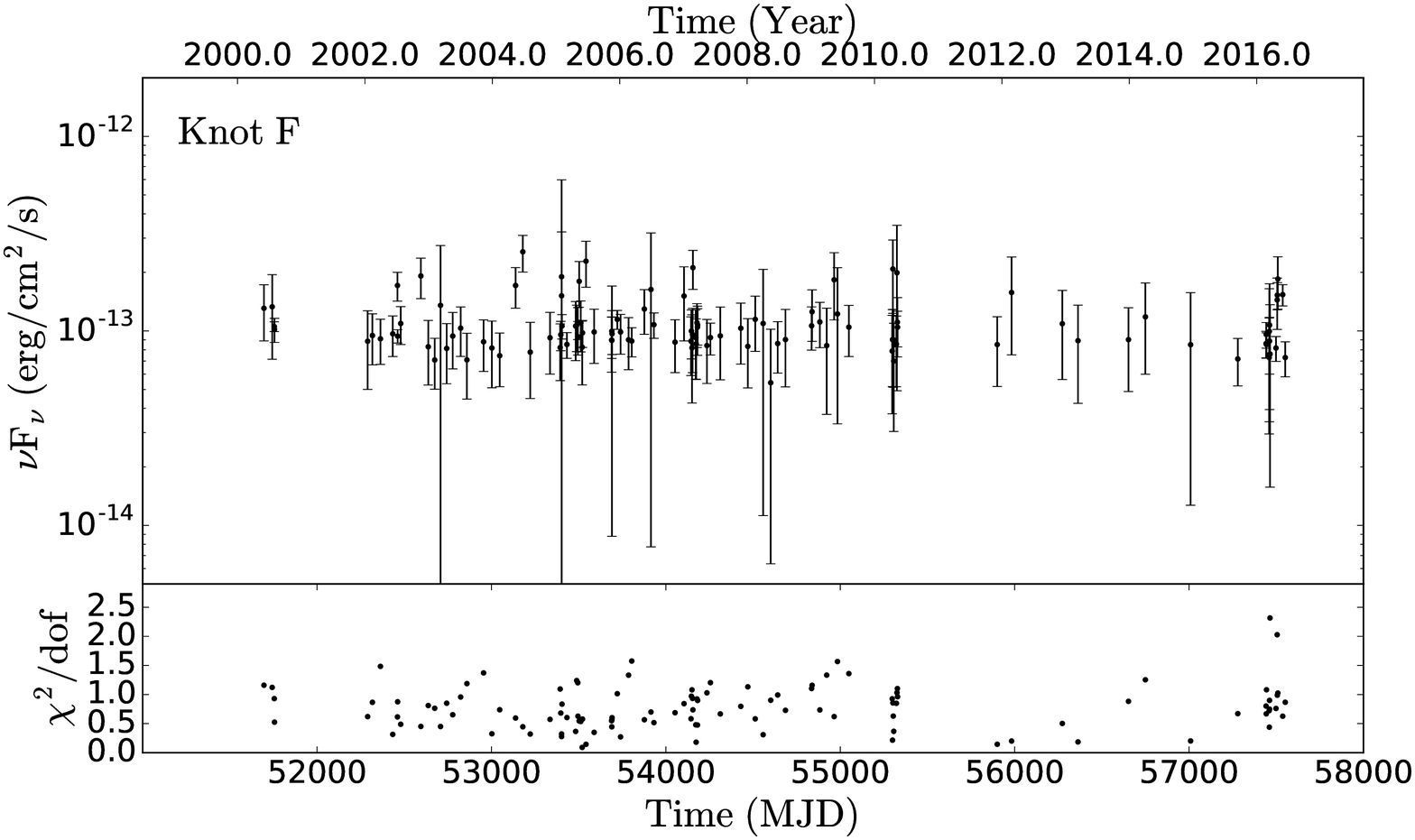}
\includegraphics[angle=0,scale=0.25]{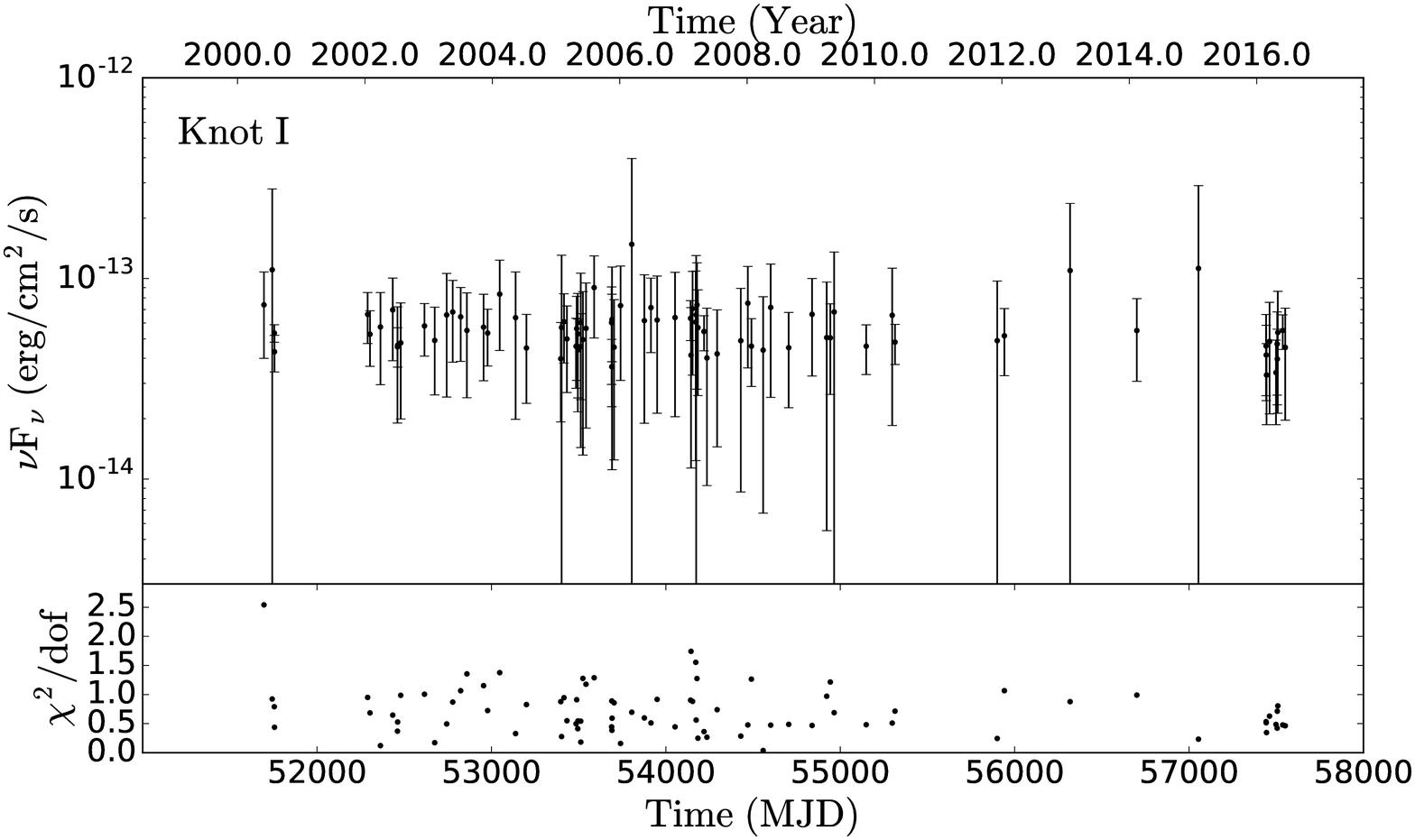}
\includegraphics[angle=0,scale=0.25]{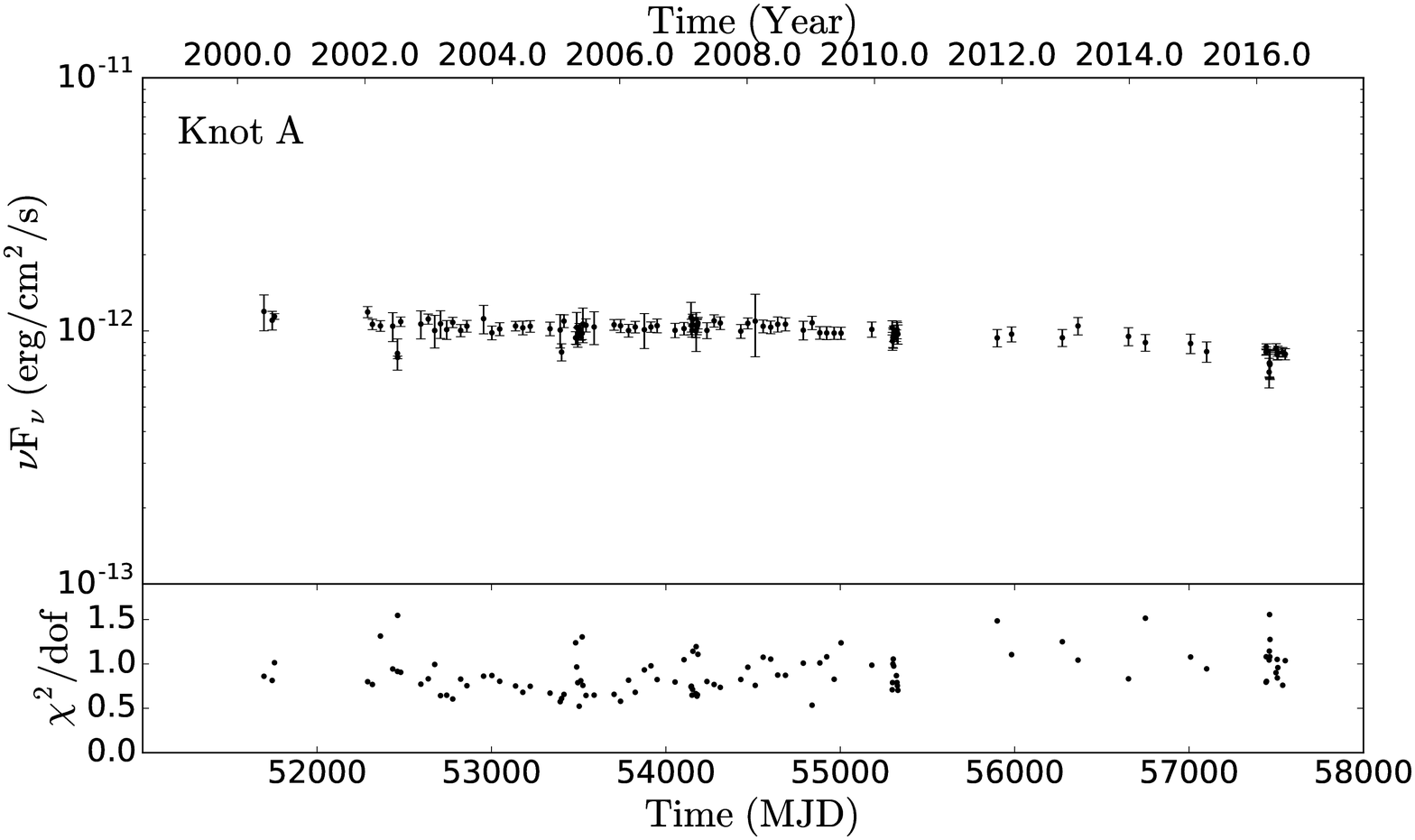}
\includegraphics[angle=0,scale=0.25]{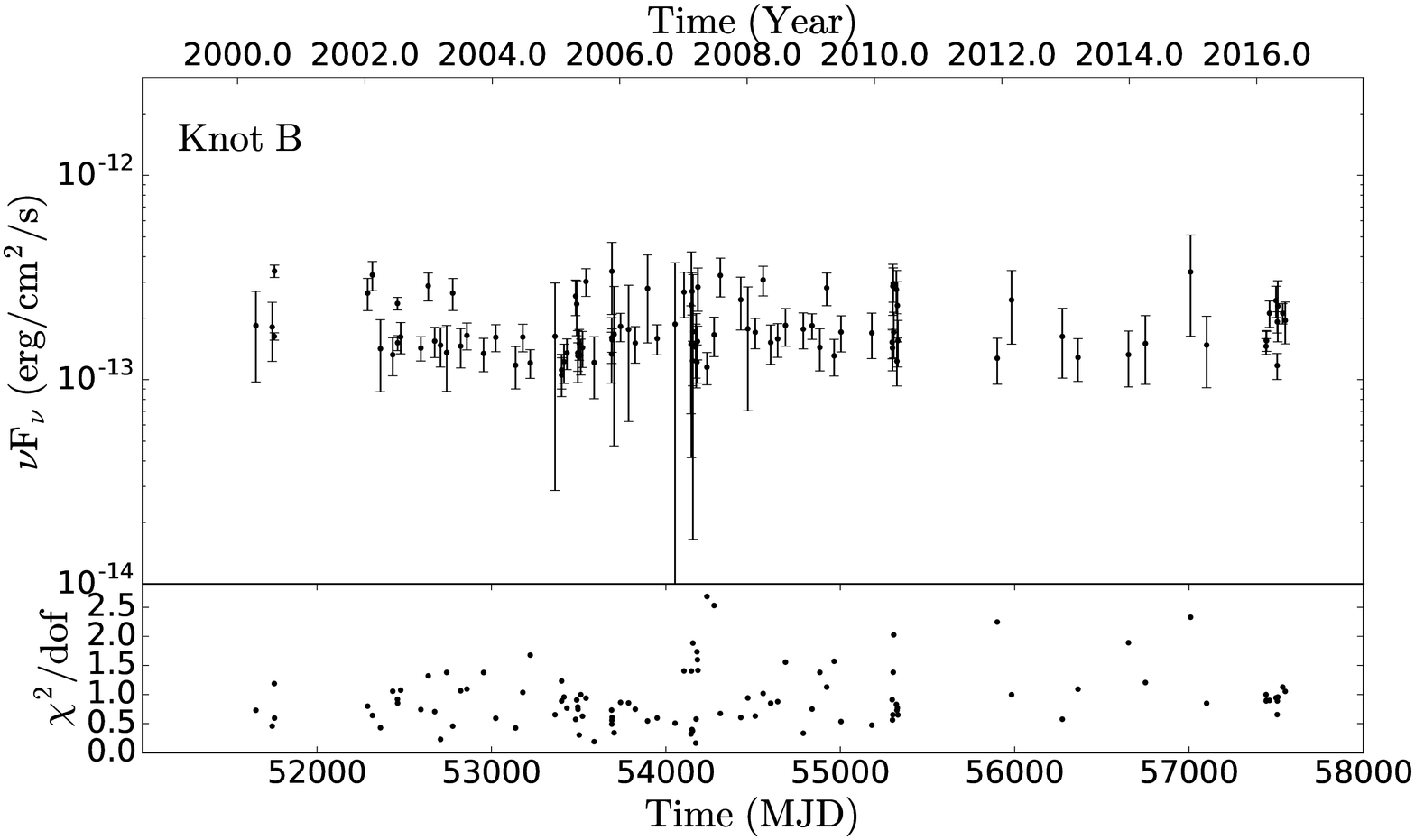}
\includegraphics[angle=0,scale=0.25]{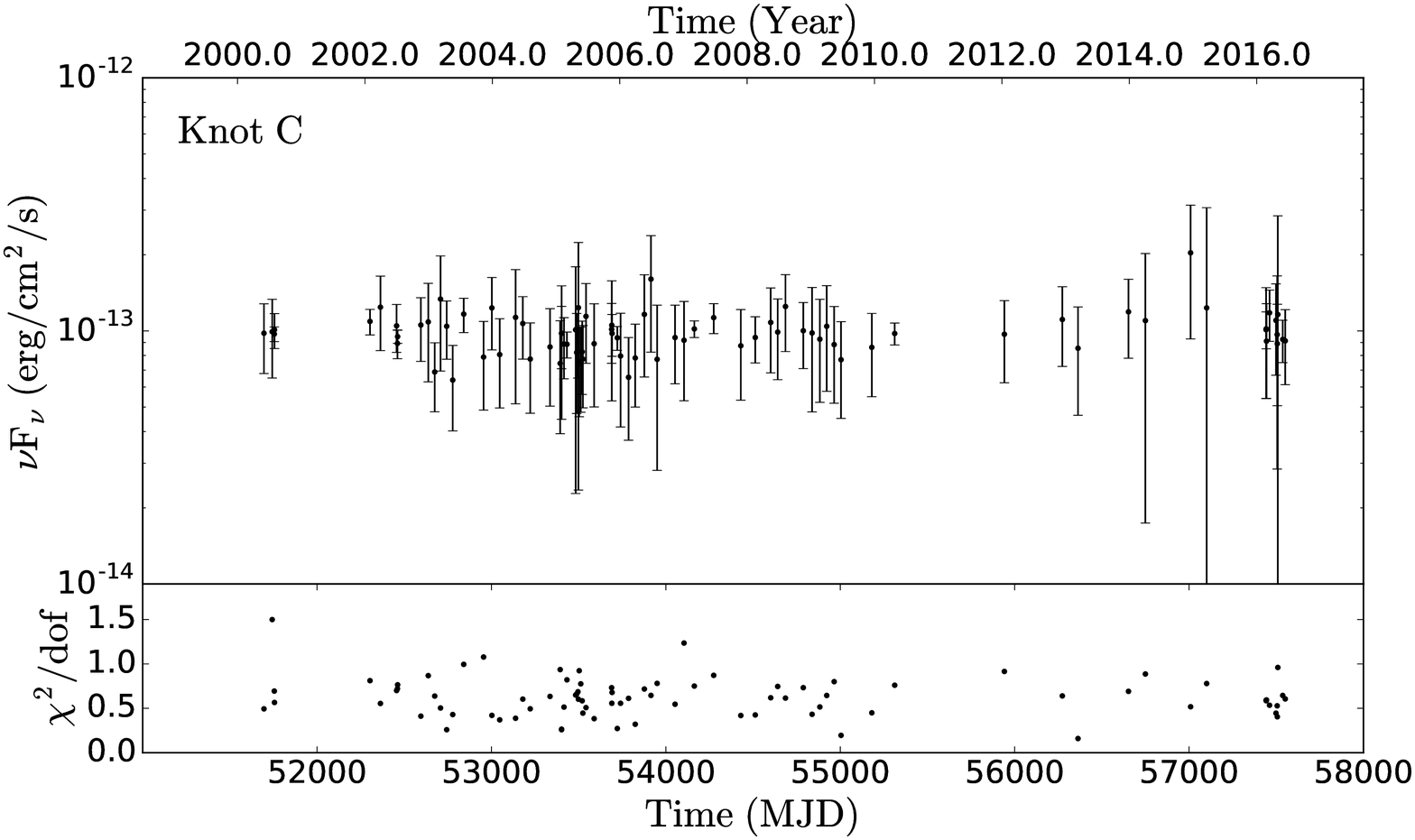}
\caption{\chandra 0.3--7 keV light curves for  knots D, E, F, I, A, B, and C during the same time interval as in Figure~\ref{fig:lc_nuc}.}
\label{fig:lc}
\end{figure*}

\subsection{Spectral analysis}\label{sec:spe_analy}
For the spectral analysis we performed an aperture photometry using {\it specextract} on the nucleus and each knot. The source and background
regions are as defined  in Figure~\ref{fig:countsmap}. 

\subsubsection{Nucleus and HST-1 during low states and in early 2010}
The nucleus is highly variable and suffers from pile-up even for observations with frame times of 0.4 s during the flare peak period H2, H3, and H4, 
which could affect the inherent spectral energy distribution (e.g. lead to spectral hardening). Thus, we only use the observations during the period of 
H1, H5, and L for temporal spectral analyses of nucleus and knot HST-1. To improve the statistics, we combine the observations on time ranges of 
a few months up to one and a half years. We simultaneously fit the spectra in each time range using a single power law plus Galactic absorption model. 
To suitably deal with the pile-up we add an additional pile-up model {\it jdpileup} to the spectral fit. The time bins and the fitted indices are listed in 
Table~\ref{table:photon_index_yearbyyear}. No obvious spectral variability seems apparent.
In addition, we also provide a detailed analysis of those observations contemporaneous to the 2010 TeV flare. In 2010 HST-1 was back to a state 
comparable with the pre-2004 time, during which pile-up was found to be fairly mild \citep{Harris06,Russell15}.
The fitted indices for the nucleus in April/May 2010 are shown in Table~\ref{table:photon_index} along with those for HST-1. 
The photon indices for the nucleus are on average close to $\Gamma \sim 2.1-2.2$, which seems compatible with earlier (non-flaring) results 
reported in \cite{Wilson02} and \cite{Perlman05}, see  also the results in Table~\ref{table:photon_index_yearbyyear}. A hint for spectral variations 
during 2010 might be seen in the nucleus, while the HST-1 spectrum appears stable, though the short exposures limit inferences. Combing 
observations for HST-1 (in a set of three) to improve statistics, however, yields results that are
compatible with no spectral change. Our values for the nucleus are similar to those in \citet{harris11}. Our results for HST-1, however, are 
harder than those reported in \citet{harris11}, yet still compatible within the $1\sigma$ error bars. We note that in our analysis the error bars are 
much larger due to the additional free parameters in the pile-up model that has been included in our analysis. For sources closer than one arcsec, 
such as the nucleus and HST-1 (which are separated by 0.86\arcsec), an `eat-thy-neighbour' effect may in principle occur \citep{Harris06,Harris09}, 
where photons arriving within the same frame time and 3x3 pixel grid are registered as a single event at the location of the pixel with the larger energy. 
As suggested by \citet{Harris09} this effect is  not expected to cause serious problems below a detector-based intensity limit of 4 keV/s 
(e.g. for the pre-2004 and post-2006 time for HST-1).

The \xray photon indices for the nucleus are in principle in the range of those achievable by Comptonisation in a hot accretion flow. Early models 
in fact assumed that the quiescent nuclear \xray emission in M87 is produced by an ADAF \citep[e.g.][]{Reynolds96,DiMatteo03}. The similarities 
of the nuclear spectrum to that of the jet and its knots, and the strong brightness increase towards the nucleus (see Figure~\ref{fig:fluxindex_distance}), 
however, suggest that this `nuclear' emission is instead dominated by the jet \citep{Wilson02}. This is supported by the fact that the luminosity of a 
hot accretion flow at low accretion rates $\dot{M}$ roughly scales with $\dot{M}^2$, i.e. $L_{ADAF} \propto \dot{M}^2$, while that of the jet with $L_j
\propto \dot{M}$, so that for low $\dot{M}$ the jet starts to dominate \citep{Yuan14,Feng17}. The evidence for strong nuclear activity over time fits 
well into this.  

\begin{table*}
\caption{\xray photon indices for the nucleus and HST-1 during the period of H1, H5, and L.}
\centering
\begin{tabular}{ccc}
\hline\hline
Time range (YYY-MM-DD)&$\Gamma_{\rm Nucleus}$&$\Gamma_{\rm HST-1}$ \\
\hline 
2002-01-16$\--$2003-08-08 &2.12$\pm$0.05&2.25$\pm$0.05 \\[0.1cm]
2008-11-17$\--$2010-05-14 &2.25$\pm$0.05&2.38$\pm$0.07 \\[0.1cm]
2011-12-04$\--$2013-03-12 &2.27$\pm$0.12&2.52$^{+0.48}_{-0.28}$ \\[0.1cm]
2013-12-26$\--$2015-03-19 &2.05$\pm$0.13&2.43$^{+0.30}_{-0.15}$ \\[0.1cm]
2016-02-23$\--$2016-06-12 &2.25$\pm$0.10&2.29$^{+0.17}_{-0.10}$ \\[0.1cm]
\hline
\end{tabular}
\tablefoot{All errors are at a 90\% confidence level.}
\label{table:photon_index_yearbyyear}
\end{table*}

\begin{table*}
\caption{\xray photon indices for the nucleus and HST-1 during the 2010 TeV flare.}
\centering
\begin{tabular}{ccc}
\hline\hline
ObsID&$\Gamma_{\rm Nucleus}$&$\Gamma_{\rm HST-1}$ \\
\hline 
11512 &1.97$\pm$0.1&2.17$^{+0.24}_{-0.21}$ \\[0.1cm]
11513 &2.14$^{+0.15}_{-0.14}$&2.20$^{+0.27}_{-0.12}$ \\[0.1cm]
11514 &1.98$^{+0.16}_{-0.15}$&2.33$^{+0.31}_{-0.16}$ \\[0.1cm]
11515 &2.10$^{+0.17}_{-0.16}$&2.23$^{+0.18}_{-0.11}$ \\[0.1cm]
11516 &1.84$^{+0.16}_{-0.13}$&2.24$^{+0.26}_{-0.13}$ \\[0.1cm]
11517 &2.26$^{+0.16}_{-0.15}$&2.23$^{+0.28}_{-0.12}$ \\[0.1cm]
11518 &2.16$^{+0.20}_{-0.18}$&2.28$^{+0.18}_{-0.14}$ \\[0.1cm]
11519 &2.19$^{+0.19}_{-0.14}$&2.22$^{+0.21}_{-0.11}$ \\[0.1cm]
11520 &2.24$^{+0.21}_{-0.18}$&2.33$^{+0.29}_{-0.11}$ \\[0.1cm]
\hline
\end{tabular}
\tablefoot{All errors are at a  90\% confidence level.}
\label{table:photon_index}
\end{table*}

\subsubsection{Combined \xray spectra of the knots}
For all knots, except HST-1, we fit multiple observations simultaneously.  For HST-1 we exclude observations during $\rm H_{2}$, 
$\rm H_{3}$, and $\rm H_{4}$ in the spectral analysis due to significant pile-up.

In the \xray band all knots can in principle be well fitted with a single power law plus Galactic absorption model, none seems to
require an obvious break or additional component within the \xray band itself (but see below). Table~\ref{table:spefits} lists the 
best-fit parameter values and the reduced~$\chi^2$ value. The reduced~$\chi^2$ values reveal that all fits are acceptable at a 
90\% confidence level. Leaving the absorption column density $N_{\rm H}$ free in the fit yields values deviating significantly from 
the Galactic one \citep{Perlman05}. \citet{Harris06} have reported a mutual dependence between $N_H$ and $\alpha$ in
the fitting process, leading to some uncertainty. For comparison, we thus also provide results with $N_{\rm H}$ frozen to the Galactic 
value. We note that we do not find evidence for a significant deviation \citep{Osone17} from a pure power law in the \xray band for 
knot A if $N_{\rm H}$ is kept frozen.

\begin{figure*}
\centering
\includegraphics[angle=0,scale=0.45]{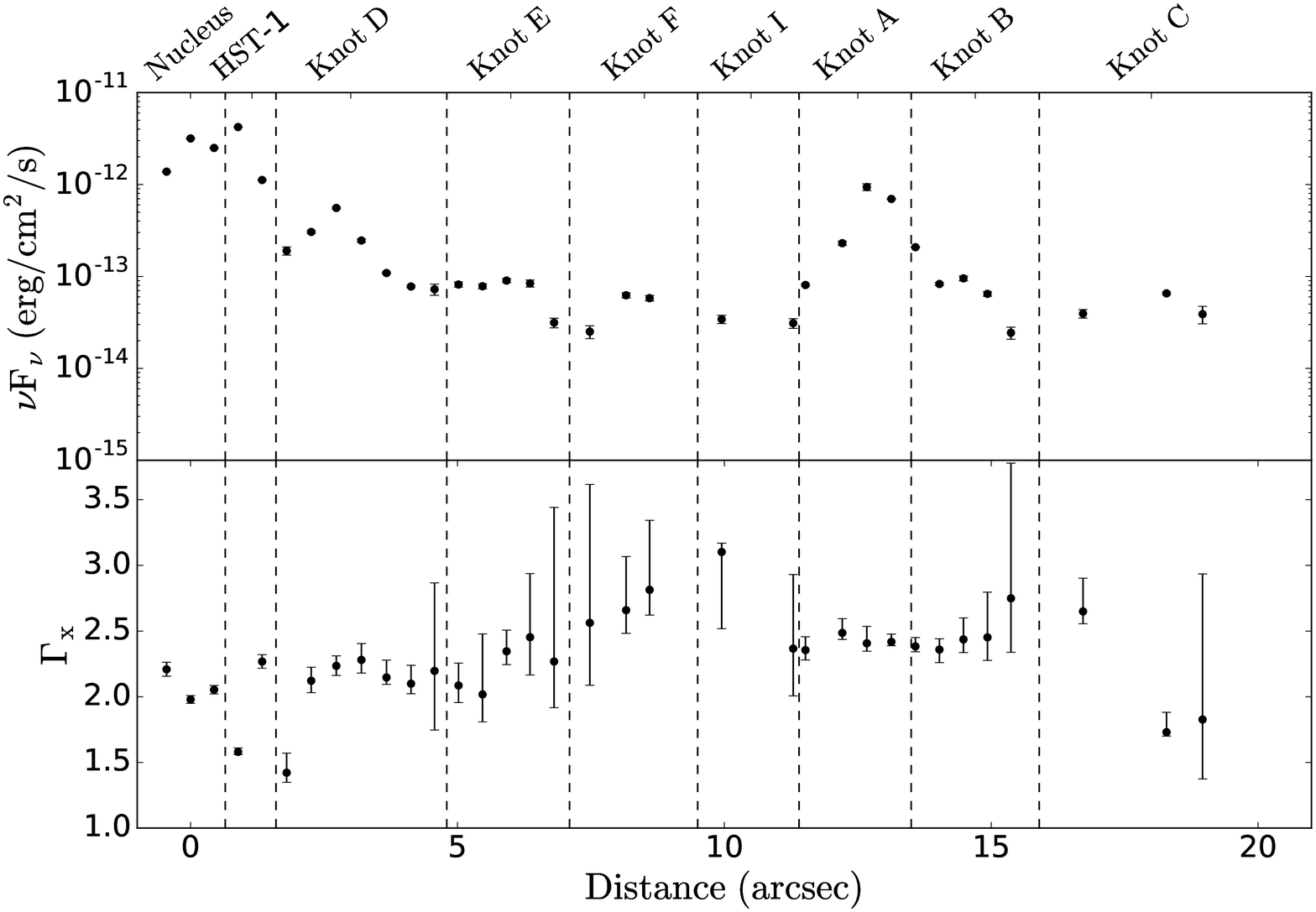}
\caption{Flux density (top panel) and power-law photon index (bottom panel) in the \xray 0.3--7 keV band along the jet. The source regions are 
separated by the dashed lines. For this representation all observations with frame time 0.4~s are used for the nucleus, while for HST-1 
only `HST-1 L' data are employed.}
\label{fig:fluxindex_distance}
\end{figure*}

The derived photon indices for different knots are significantly different, ranging from $\simeq1.89$ (knot C) to $\simeq2.61$ (knot F). 
The results roughly match those obtained by \citet{Wilson02,Wilson2004} and \citet{Perlman05}. 
In Figure~\ref{fig:fluxindex_distance} we plot the profiles of the flux density and photon index $\alpha_{\rm x}$ along the jet. For 
the regions of the nucleus,  knots HST-1, D, and A, we define the size of the stripes of 0.45" and 2.0" along and perpendicular to the jet.
For the other regions, we use $0.45\arcsec \times 1.5\arcsec$ rectangular regions correspondingly. If the signals are too weak, we combine
several strips together. We again used the single power law plus Galactic absorption model to fit each region. Finally, we used the best-fit  
model to simulate the energy flux of the de-convolved non-thermal emission. In this process we divided the \xray energy interval between 
0.3 keV and 7 keV into four bins in logarithmic space. The derived \xray data points are listed in Table \ref{table:flux_xray}. With the exemption
of the outermost knot C, the \xray photon index along the jet exhibits a trend similar to that reported in \citet{Perlman05}, with slight but 
significant index variations ranging from $\simeq 2.2$ (e.g. in knot D) to $\simeq 2.4-2.6$ (in knots F, A, and B). There is little evidence
that the inter-knot regions  have significantly steeper spectra than the adjacent knots,  reinforcing the need for a distributed acceleration 
mechanism.

%
%
%
%
According to Figure~\ref{fig:fluxindex_distance} the X-ray flux density along the jet exhibits significant changes on comparably small 
scales around the knot D and knot A regions. Radio and optical observations indicate that the knot D region is in fact composed 
of several subregions \citep[e.g.][see also below]{Perlman05,Avachat16}. Our modelling of the broad-band SED of the knots in Sect.~\ref{sec:fitting}, 
on the other hand, suggests that for  knot regions B and C, and to some (weaker) extent also for  knot region D, an additional X-ray 
emitting component is needed to achieve a satisfactory SED fit. It seems conceivable that these X-ray features are related to a blending of 
unresolved longitudinal or radial substructures (i.e. related to lateral jet stratification).

%


\section{SED modelling}\label{sec:fitting}
\begin{figure*}
\includegraphics[angle=0,scale=0.36]{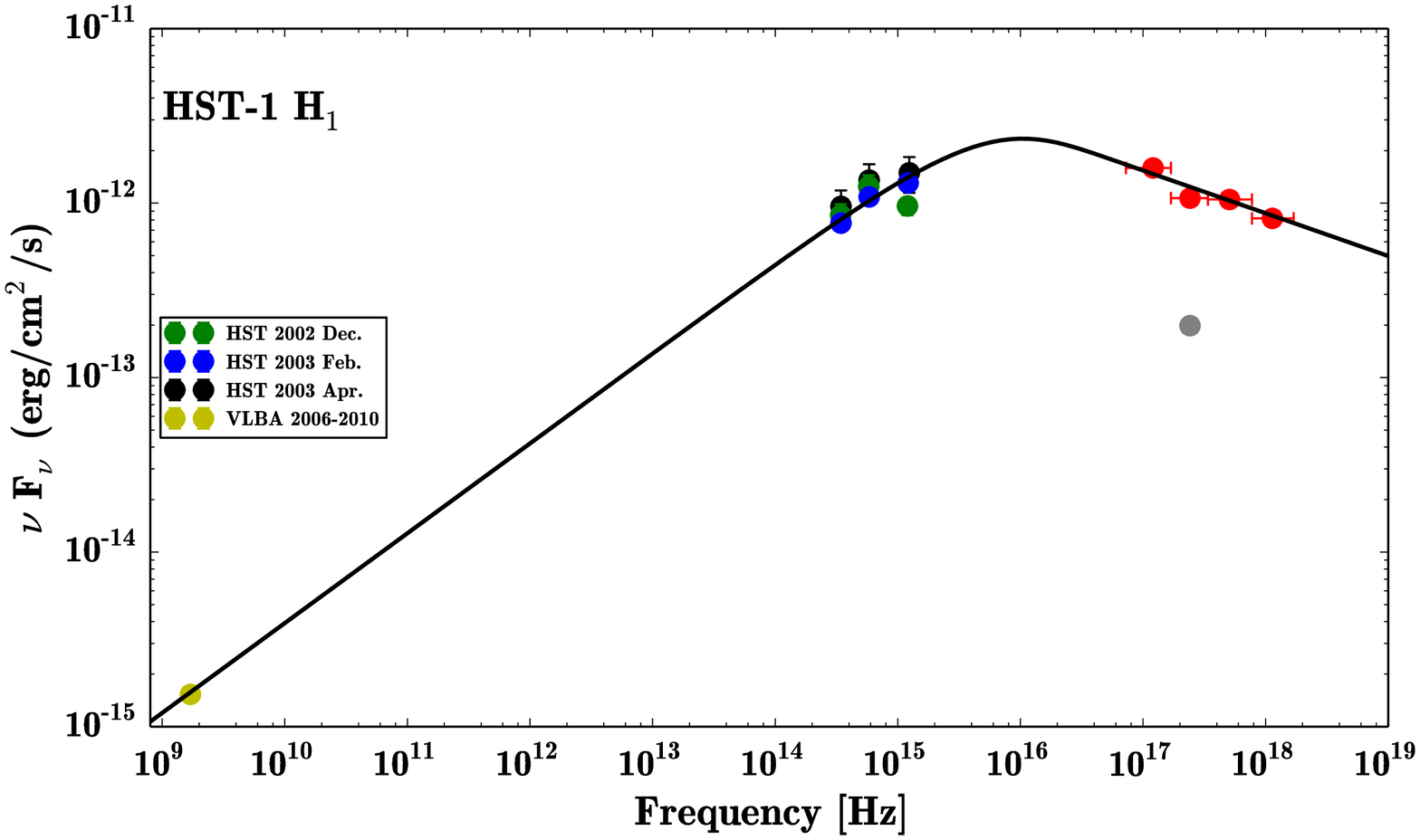}
\includegraphics[angle=0,scale=0.36]{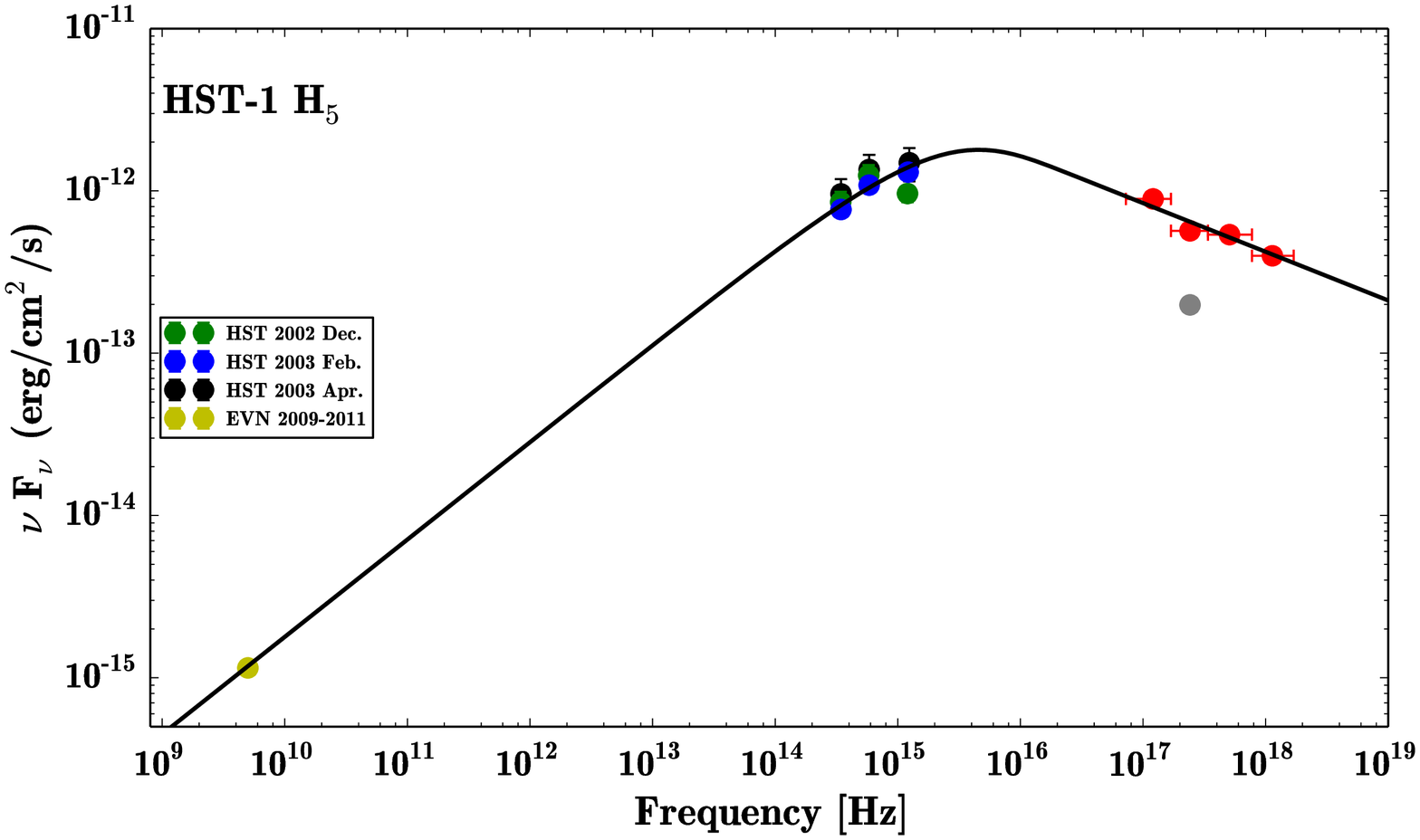}
\includegraphics[angle=0,scale=0.36]{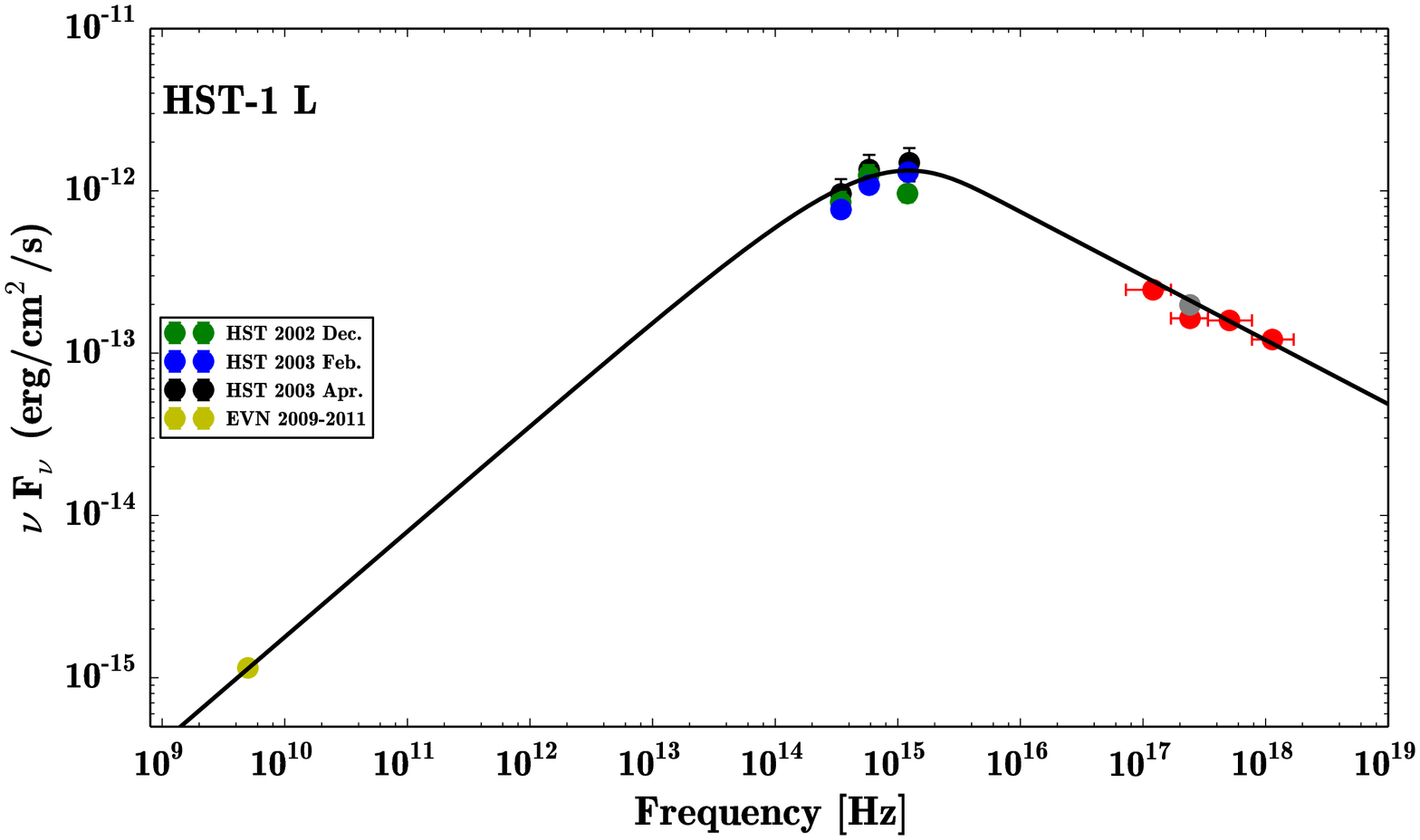}
\includegraphics[angle=0,scale=0.36]{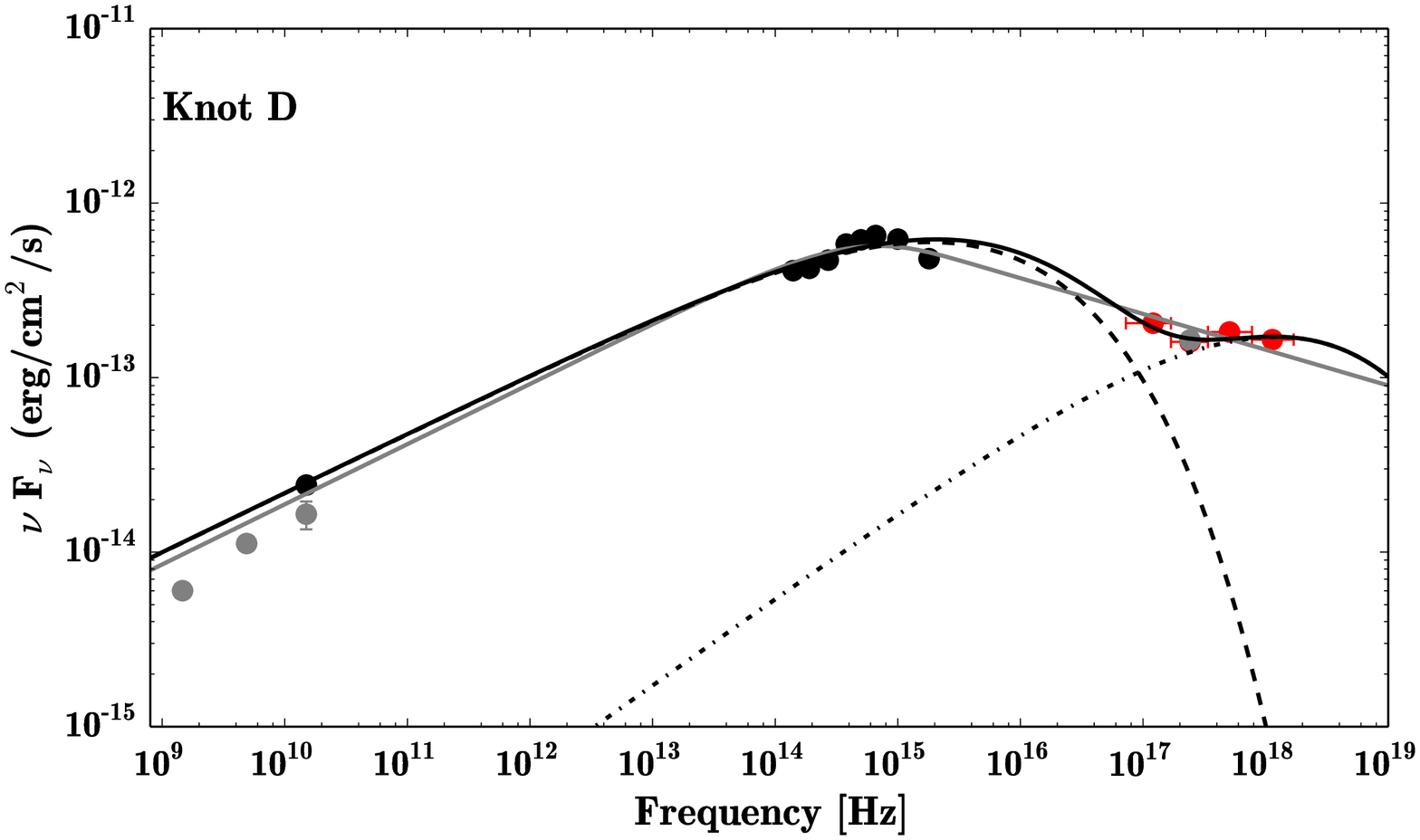}
\includegraphics[angle=0,scale=0.36]{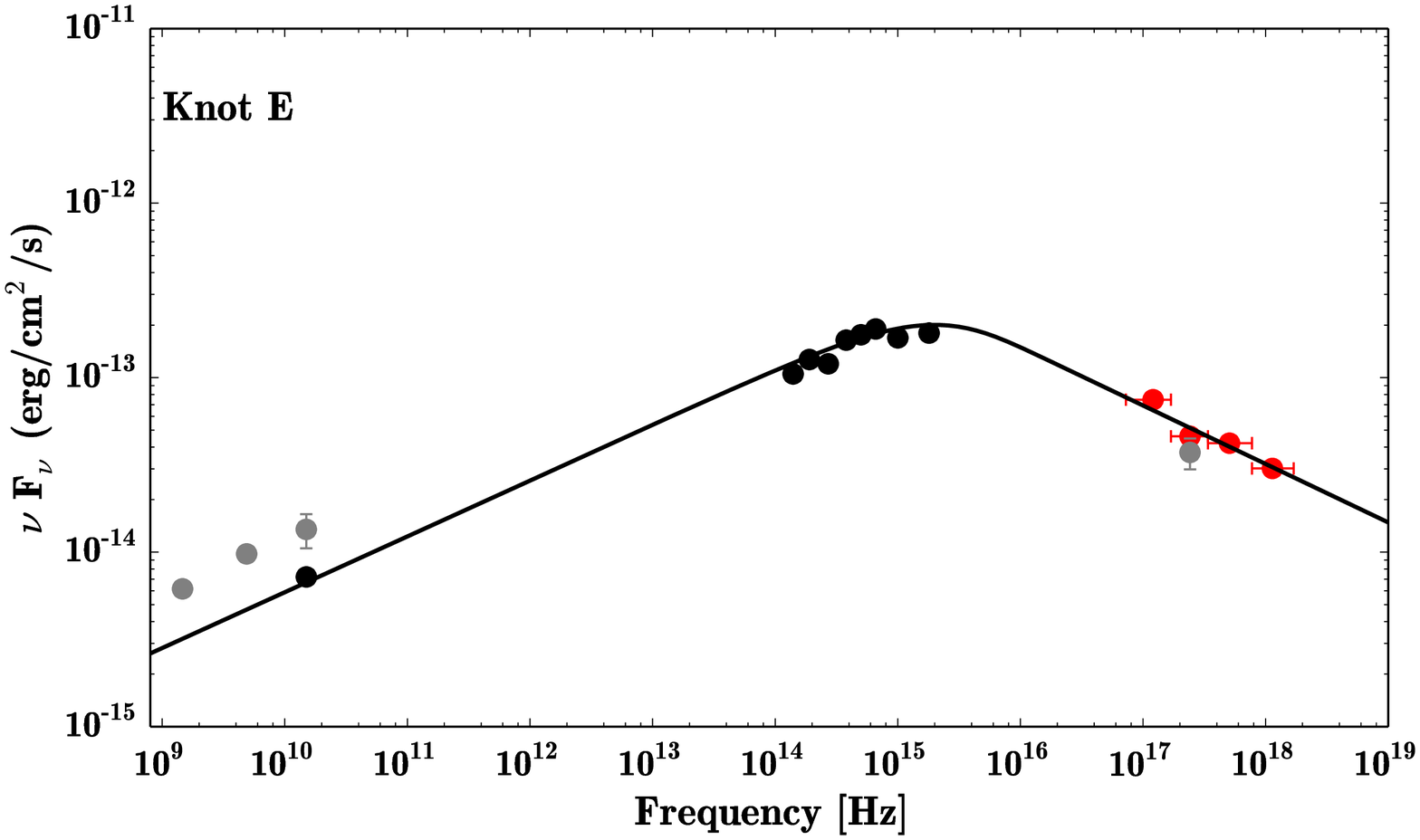}
\includegraphics[angle=0,scale=0.36]{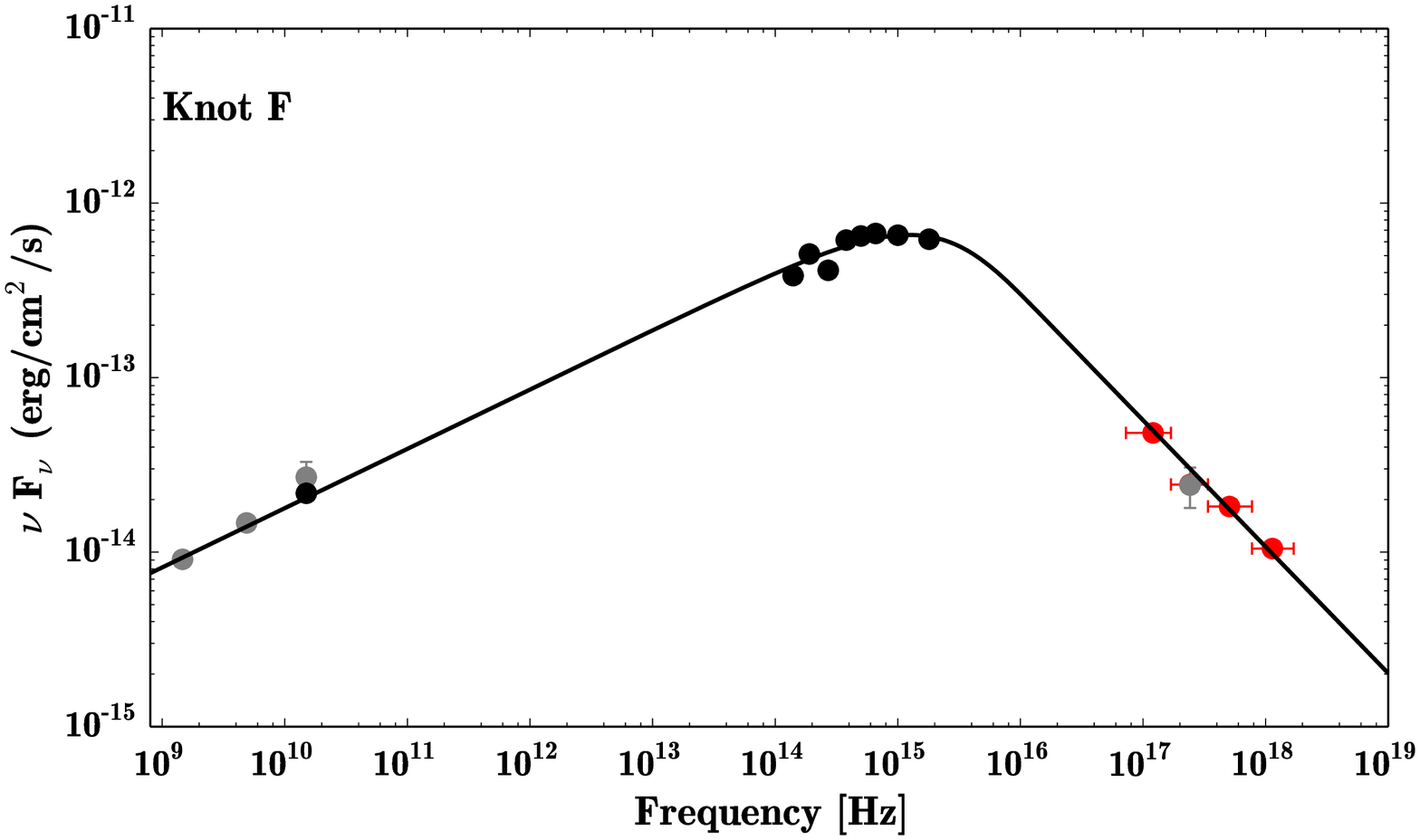}
\includegraphics[angle=0,scale=0.36]{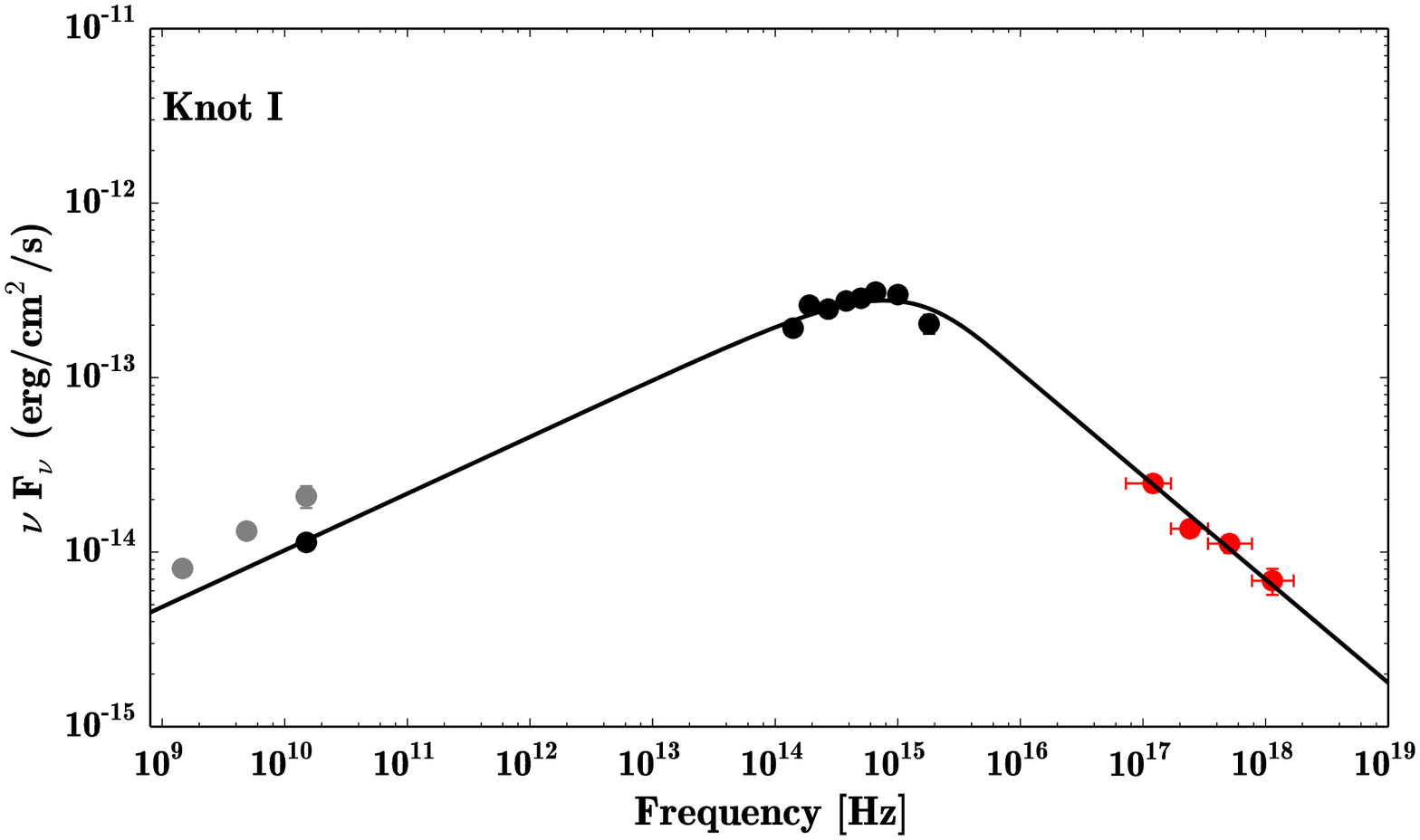}
\includegraphics[angle=0,scale=0.36]{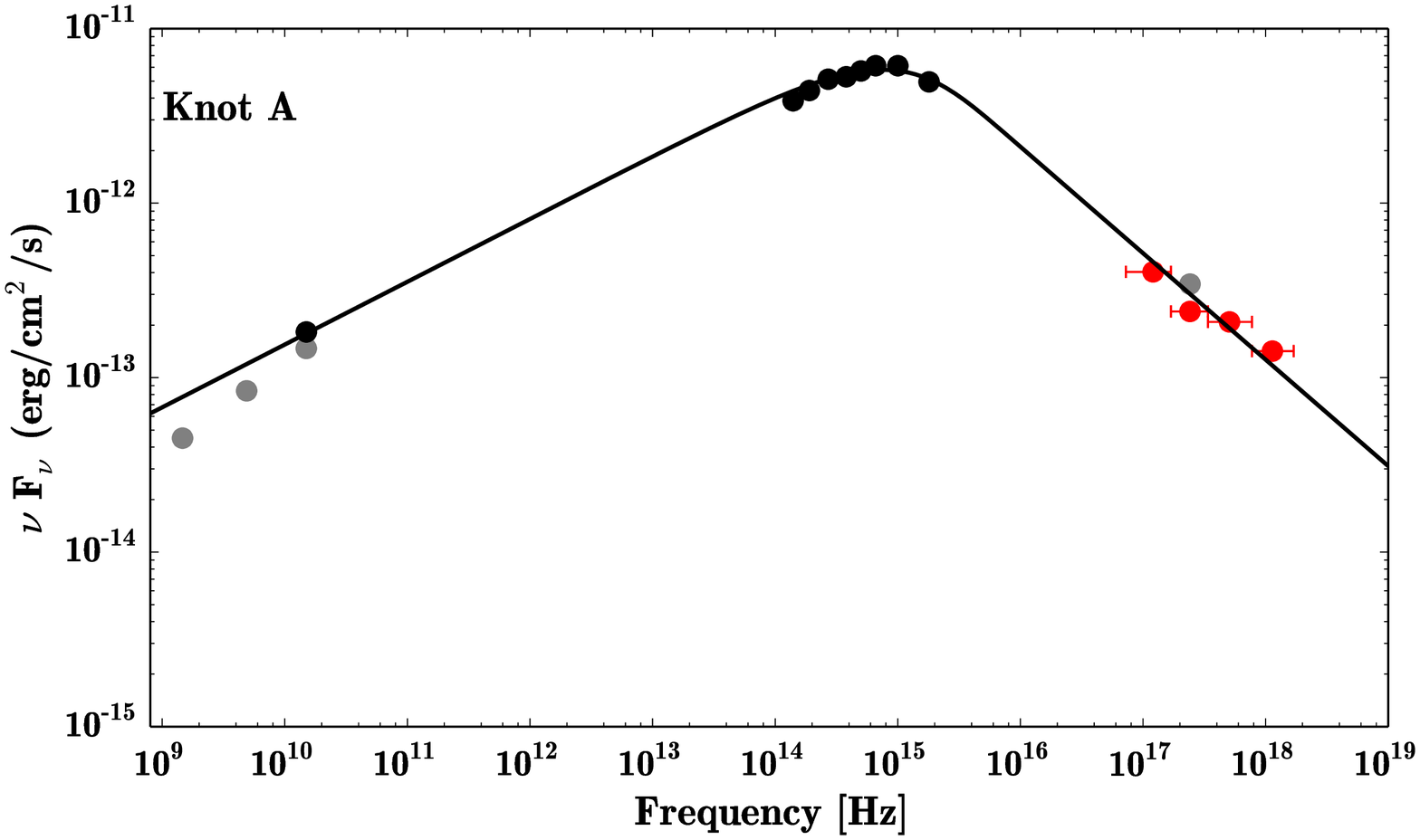}
\caption{SEDs fitting results assuming a synchrotron origin of the \xray emission. The curves represent the best-fit models. For  knots D, 
B, and C both a broken power-law fit and a fit based on two exponential cut-off power-law components (dashed and dot-dashed curves, with 
superposition corresponding to the solid curve) assuming $\beta$ = 1 are shown. The \xray flux densities from \citet{Marshall02} are also 
plotted for comparison (grey symbols). } 
\label{fig:sed}
\end{figure*}

In order to gain further insights, we constructed the radio to \xray SEDs of the knots and provided model fits to infer the characteristics of the 
parent particle distribution. We selected contemporaneous multiwavelength  data to ensure an adequate reconstruction. The optical to \xray 
spectra of the $\rm H_{1}$, $\rm H_{5}$, and L state of knot HST-1 are taken from the 2002 December, 2003 February, and 2003 April HST 
observations in \citet{Perlman03}, respectively. The radio data are from \citet{Giroletti12} based on observations between 2006 and 2010 
with the Very Long Baseline Array (VLBA) at 1.7 GHz for $\rm H_{1}$, and between 2009 and 2011 with the European VLBI Network (EVN) 
at 5 GHz for the statuses $\rm H_{5}$ and L.
We also used the published optical--near-infrared data from \citet{Perlman01}, and the ultraviolet data from \citet{Waters05} for  knots 
D, E, F, I, A, B, and C. The black points in the radio energy band are from \citet{Perlman01}. We did not include the radio data points of 
\citet{Biretta91} in the spectral fitting, yet regarded them as a reference for the radio spectral indices because of the uncertainties with
regard to these measurements \citep[see e.g.][]{Perlman05}. 
It should also be noted that although our region C could be further divided into the subregions C1 and C2, we only include the ultraviolet  
fluxes for C1. This simplification appears justified considering the much lower ultraviolet flux from C2.  For comparison, we also plot the 
\xray flux density from \citet{Marshall02} with grey symbols in Figure~\ref{fig:sed}.

%

There are in principle several possible non-thermal scenarios for the origin of large-scale \xray emission, such as inverse Compton  
(IC) upscattering of either synchrotron (synchrotron self-Compton; IC-SSC) or cosmic microwave background (IC-CMB) photons, or direct synchrotron emission 
of relativistic electrons or protons \citep[e.g.][]{Aharonian02,Harris02,Zhang09, Zhang10, Georganopoulos16}. As discussed in \citet{Wilson02}, to account for the keV 
\xray flux by means of leptonic IC-SSC processes, the magnetic field has to be substantially lower than equipartition. In the case of IC-CMB it is 
possible to boost the \xray flux by relativistic bulk motion, but this requires high Doppler factors $\delta \simeq (10-40)$ and very small angles $\theta 
\simeq (1.2^\circ-4.7^\circ)$ between the jet and the line of sight \citep{Harris02}. Though HST observations indicate that moderate superluminal 
motion in M87 could persist out to knot C \citep{Meyer13}, such  values remain highly unlikely. Moreover, IC radiation by a power-law electron 
distribution in the Thomson regime is expected to exhibit a spectral index similar to that for synchrotron radiation, yet the observed \xray spectra 
of the knots are all significantly steeper than the radio spectra \citep[exhibiting spectral indices $\alpha_{rr} \sim 0.5$, e.g.][]{Biretta91}. An IC origin 
of the \xray emission in M87 thus appears disfavoured \citep[see also][]{Biretta91,Meisenheimer96,Wilson02,Perlman05, Zhang10}, and we therefore only 
consider an electron synchrotron origin in the following.


\begin{table*}
\caption{Results for the \xray spectral fitting of the knots.}
\centering
\begin{tabular}{ccccccc}
\hline\hline
Component & $\Gamma$\tablefootmark{a} & \nh \tablefootmark{b} &$K$\tablefootmark{c}&$\nu \rm F_\nu$\tablefootmark{d} & $\rm reduced~\chi^2$ & dof \\

 & &($\rm \times 10^{20}/cm^{2}$) &($\rm \times 10^{-5}photons/keV/s$)&($\rm \times 10^{-13}erg/cm^{2}/s$) && \\ [0.1cm]
\hline
HST-1 $\rm H_{1}$&2.25$\pm$0.05&4.74$^{+0.78}_{-0.75}$&94.88$^{+3.20}_{-3.10}$&44.73$\pm$0.62&0.98&2174\\[0.1cm]
HST-1 $\rm H_{5}$&2.31$\pm$0.06&1.21$^{+1.15}_{-1.10}$&50.59$^{+2.45}_{-2.16}$&23.62$\pm$0.59&0.93&1293\\[0.1cm]
HST-1 L&2.27$^{+0.06}_{-0.04}$&<1.21&14.53$^{+0.69}_{-0.35}$&6.86$\pm$0.15&1.08&1801\\[0.1cm]
D&2.22$\pm$0.03&2.50$\pm$0.5&17.11$\pm$0.4&8.13$\pm$0.10&0.72&5526\\[0.1cm]
E&2.35$\pm$0.05&2.34$^{+0.86}_{-0.83}$&4.10$^{+0.16}_{-0.15}$&1.91$\pm$0.04&0.70&2726\\[0.1cm]
F&2.61$^{+0.07}_{-0.02}$&<0.46&2.15$^{+0.06}_{-0.05}$&1.00$\pm$0.01&0.74&1720\\[0.1cm]
I&2.50$^{+0.04}_{-0.11}$&<0.57&1.20$^{+0.02}_{-0.08}$&0.56$\pm$0.04&1.02&576\\[0.1cm]
A&2.41$\pm$0.02&<0.10&21.22$^{+0.13}_{-0.36}$&9.85$\pm$0.02&0.90&4811\\[0.1cm]
B&2.35$^{+0.06}_{-0.02}$&<0.8&3.49$^{+0.13}_{-0.06}$&1.63$\pm$0.02&0.89&2480\\[0.1cm]
C&1.89$^{+0.09}_{-0.03}$&<0.85&1.78$^{+0.12}_{-0.04}$&0.94$\pm$0.02&0.64&2145\\[0.1cm]
\hline
\hline
F&2.79$\pm$0.04&2.4 (frozen)&2.36$\pm$0.05&1.13$\pm$0.02&0.75&1721\\[0.1cm]
I&2.60$\pm$0.04&2.4 (frozen)&1.28$\pm$0.03&0.60$\pm$0.01&1.04&577\\[0.1cm]
A&2.55$\pm$0.01&2.4 (frozen)&23.10$\pm$0.2&10.74$\pm$0.05&0.96&4812\\[0.1cm]
B&2.49$\pm$0.03&2.4 (frozen)&3.85$\pm$0.06&1.79$\pm$0.02&0.90&2481\\[0.1cm]
C&2.01$\pm$0.04&2.4 (frozen)&1.99$\pm$0.05&1.0$\pm$0.02&0.65&2146\\[0.1cm]
\hline

\end{tabular}
\tablefoot{All errors are at a  90\% confidence level. $\rm H_{1}$, $\rm H_{5}$, and L are defined as shown in Figure \ref{fig:lc_hst}.
\tablefoottext{a}{Photon index.}
\tablefoottext{b}{Equivalent hydrogen-absorbing column density.}
\tablefoottext{c}{Amplitude of power-law model.}
\tablefoottext{d}{Total integrated energy flux over 0.3--7 keV.}
The bottom part provides the results if $N_H$ is frozen to the Galactic value.
}
\label{table:spefits}
\end{table*}

\begin{table*}
\caption{Deconvolved \xray flux densities of the knots for different energy bands.}
\centering
\begin{tabular}{ccccc}
\hline\hline
Energy [keV] & $0.3 - 0.7$ & $0.7 - 1.4$ &$1.4 - 3.2$ &$3.2 - 7.0$ \\
\hline
Component &  & Flux density [$\rm \times 10^{-14} erg/cm^{2}/s$]  & &  \\
\hline
HST-1 $\rm H_{1}$&159.06$\pm$3.23&106.78$\pm$1.47&104.89$\pm$2.10&81.78$\pm$2.64 \\[0.1cm]
HST-1 $\rm H_{5}$&89.26$\pm$3.11&56.74$\pm$1.29&53.78$\pm$1.68&39.83$\pm$2.33 \\[0.1cm]
HST-1 L&24.58$\pm$0.97&16.40$\pm$0.40&15.92$\pm$0.66&12.14$\pm$0.97 \\[0.1cm]
D&23.03$\pm$0.33&17.22$\pm$0.17&18.71$\pm$0.24&16.10$\pm$0.33 \\[0.1cm]
E&7.47$\pm$0.20&4.61$\pm$0.10&4.21$\pm$0.12&3.02$\pm$0.14 \\[0.1cm]
F&4.82$\pm$0.11&2.44$\pm$0.03&1.83$\pm$0.04&1.05$\pm$0.04 \\[0.1cm]
I&2.48$\pm$0.26&1.36$\pm$0.09&1.12$\pm$0.13&0.69$\pm$0.12 \\[0.1cm]
A&40.38$\pm$0.19&23.94$\pm$0.06&20.86$\pm$0.09&14.19$\pm$0.12 \\[0.1cm]
B&6.33$\pm$0.10&3.93$\pm$0.05&3.58$\pm$0.10&2.56$\pm$0.05 \\[0.1cm]
C&2.25$\pm$0.04&2.00$\pm$0.03&2.59$\pm$0.06&2.67$\pm$0.10 \\[0.1cm]
\hline
\end{tabular}
\label{table:flux_xray}
\end{table*}

We use {\it Naima} \citep{naima} to explore the characteristics of the radiating particle distribution. Naima is a numerical 
package that includes a set of non-thermal radiative models and a spectral fitting procedure. The best-fit and uncertainty 
distributions of spectral model parameters are derived through Markov chain Monte Carlo \citep[MCMC;][]{Foreman13} 
sampling of their likelihood distributions. The code allows us to implement different functions and includes tools to perform 
MCMC fitting of non-thermal radiative processes to the data. Given the radio to \xray data of the knots in M87, the simplest 
functional form to take into account is a broken power law,
\begin{equation}
N(E) =
\begin{cases}
AE^{-\alpha_{\rm 1}} & \quad  E\leq E_{\rm break}\\
A E_{\rm break}^{(\alpha_{\rm 2}-\alpha_{\rm 1})}~E^{-\alpha_{\rm 2}} & \quad  E>E_{\rm break}\,.\\
\end{cases}
\label{equ:bpl}
\end{equation}
For the fitting process, the parameters  $A$, $\alpha_{1}$, $E_{\rm break}$, and $\alpha_{2}$ are left as free parameters and we 
assume non-relativistic motion for the entire jet.
Our results (Table ~\ref{table:derived_pra}) show that the SEDs of  knots E, F, I, and A could in principle be 
satisfactorily described by synchrotron emission of a broken power-law electron distribution with an index $\alpha_{\rm 1} 
\simeq 2.3$ resembling that of particle acceleration at highly relativistic shocks \citep[e.g.][]{Achterberg01}. The second
index $\alpha_2$ for these knots deviates from that of a classical cooling break. This is interesting, but could possibly be 
accounted for by deviations from a homogeneous one-zone approach and/or some additional re-acceleration \citep[e.g.][]{Liu07,
Sahayanathan08,Liu17}. Diffusive synchrotron radiation from random small-scale magnetic fields could offer an alternative 
explanation \citep{Fleishman06}, although the observed high radio and optical polarisation \citep[e.g.][]{Avachat16} may 
complicate this interpretation.
As an alternative a log-parabola model \citep[cf.][]{Massaro04,Tramacere07},
\begin{equation}
N(E) = A~\left(\frac{E}{100~\rm GeV}\right)^{-\alpha-\beta~{\rm log}~ \left(\frac{E}{100~\rm GeV}\right)},
\label{equ:logparabola}
\end{equation}
has also been employed to fit the SEDs, and the corresponding results are presented in Table~\ref{table:derived_pra}. 
A comparison of the MLL values for a broken power-law and a log-parabola model, using the Bayesian Information 
Criterion (BIC), shows that the first is preferred over the second.

The SEDs of  knots D, B, and C, on the other hand, cannot be well-fitted assuming a homogeneous source region and 
the broken power-law from Eq.~\ref{equ:bpl}. In principle, this should not come as a surprise given the morphological 
complexity of the knots and the evidence of internal jet stratification \citep[e.g.][]{Perlman05,Avachat16}. The findings are
nevertheless quite interesting and  provide a first indication that the \xray emission of large-scale AGN jets might reveal 
some excess above a simple power-law extension from the optical fluxes, and in fact consist of multiple contributions. 
To explore this in more detail, we added an additional electron component to fit the \xray data. For simplicity we chose a 
two exponential cut-off power-laws model,
\begin{equation}
N(E) = A_{1}~E^{-\alpha_{1}}~{\rm exp}\left[-\left(\frac{E}{E_{\rm cutoff1}}\right)^{\beta}\right]
          +A_{2}~E^{-\alpha_{2}}~{\rm exp}\left[-\left(\frac{E}{E_{\rm cutoff2}}\right)^{\beta}\right]\,,
\label{equ:2ecpl}
\end{equation} treating again $A_{1}$, $\alpha_{1}$, $E_{\rm cutoff1}$, $A_{2}$, $\alpha_{2}$, and $E_{\rm cutoff2}$  as 
free parameters for the fit. We note that alternative descriptions (e.g. two broken power laws with exponential cut-off) 
are in principle possible, yet require that some additional parameters be fixed. In our phenomenological fit the two components 
are treated as independent. We note, though, that the correlation between the \xray flux maxima and the synchrotron break 
frequencies \citep[e.g.][]{Perlman05} suggest that the \xray emitting particles are not completely independent from those 
at lower energies. The results should thus be treated only as a first guide. To reduce the free parameters in our fit (Eq.~\ref{equ:2ecpl}), we fixed the parameter $\beta$ to 1 and 2, corresponding to the escaping and cooling-dominated regime, 
respectively. We assume a magnetic field strength of $B = 300\ \rm \mu G$ \citep{Heinz97, Marshall02} as a reference value 
in all our calculations. This is certainly a simplification as the magnetic field strength is expected to be different for different 
regions, for example it can vary by a factor of $\sim2$ when different knots are compared \citep[e.g.][]{Biretta91,Meisenheimer96}. We 
note however, that in the fitting procedure $B$ scales as $E_{\rm cutoff}^{-2}$ and $A^{-0.5}$, so that a different $B$ only 
affects the best-fit value of the break or cut-off energy, and the absolute normalisation. 

The  derived model parameters as well as the total energy in non-thermal electrons, $W_{\rm e}$, with $1 \sigma$ errors are 
listed in Table~\ref{table:derived_pra}. Figure~\ref{fig:sed} shows the corresponding synchrotron model fits for the observed 
SEDs of the knots. The curves represent the SED models with the maximum likelihood. For   knots D, B, and C, the dashed 
curves and the dot-dashed curves are derived from two exponential cut-off power-law components, respectively, as noted above. 
The derived values for the required energy in non-thermal electrons range up to $W_{\rm e} \simeq 10^{53}$ erg (knot A) and 
imply a jet kinetic power $P_{\rm j} \gtrsim 10^{43}$ erg/s, compatible with other estimates for the general jet power (see 
Introduction). The energy density in electrons, $u_e$, is typically comparable to (or somewhat less than) that in magnetic fields, 
$u_B=B^2/(8\pi)$, ensuring efficient confinement.

\begin{figure*}
\includegraphics[width=0.5\textwidth, height=160px]{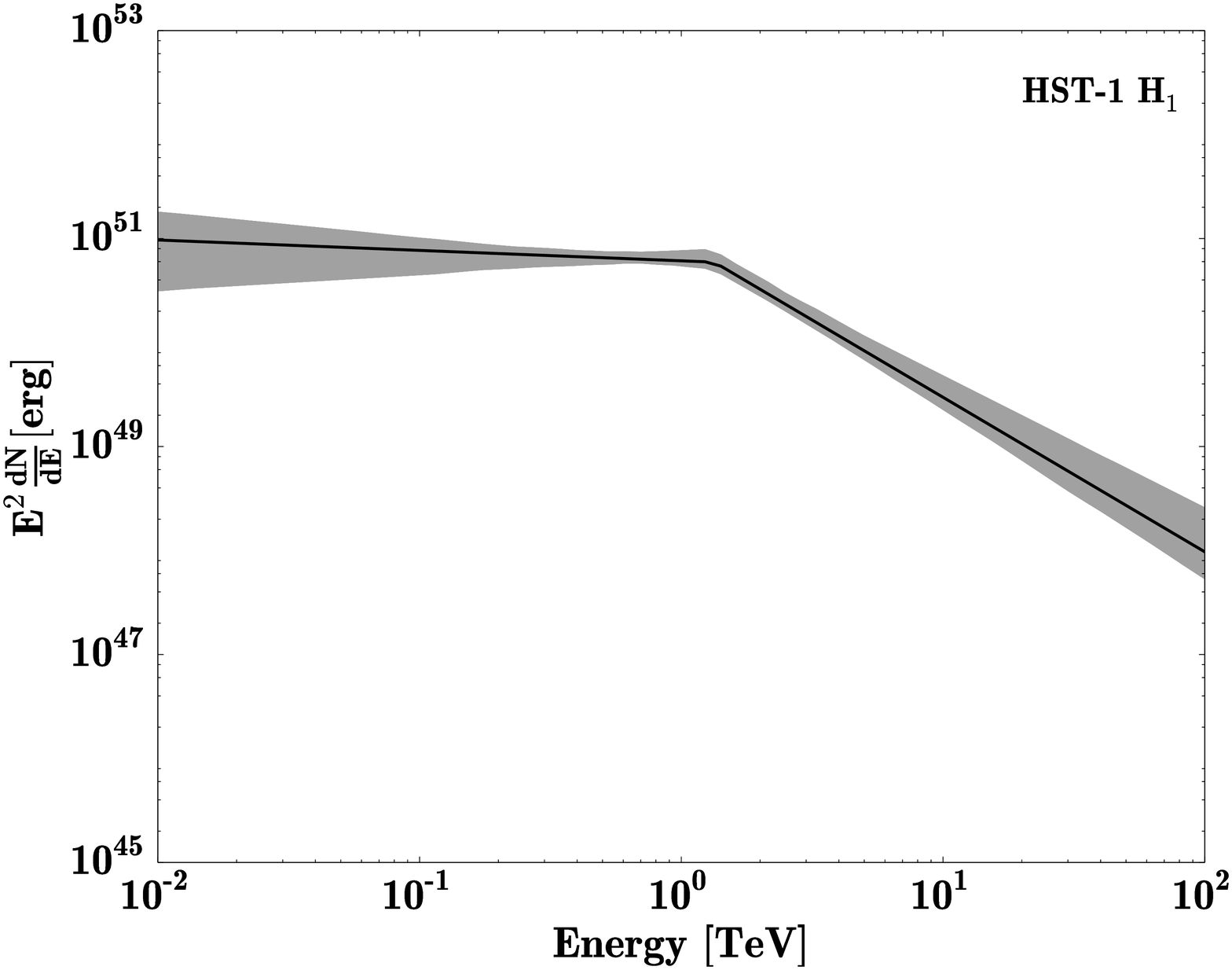}
\includegraphics[width=0.5\textwidth, height=160px]{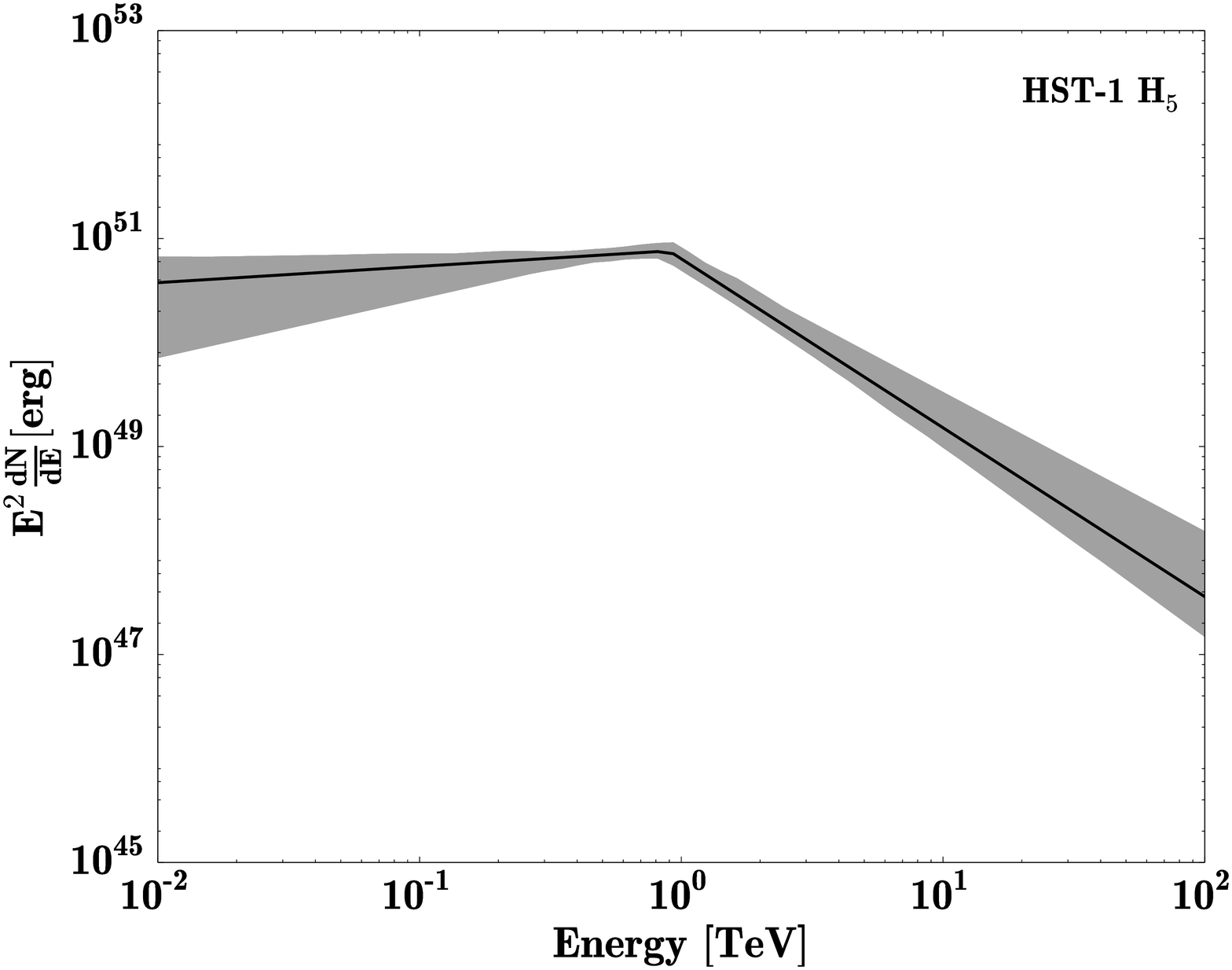}
\includegraphics[width=0.5\textwidth, height=160px]{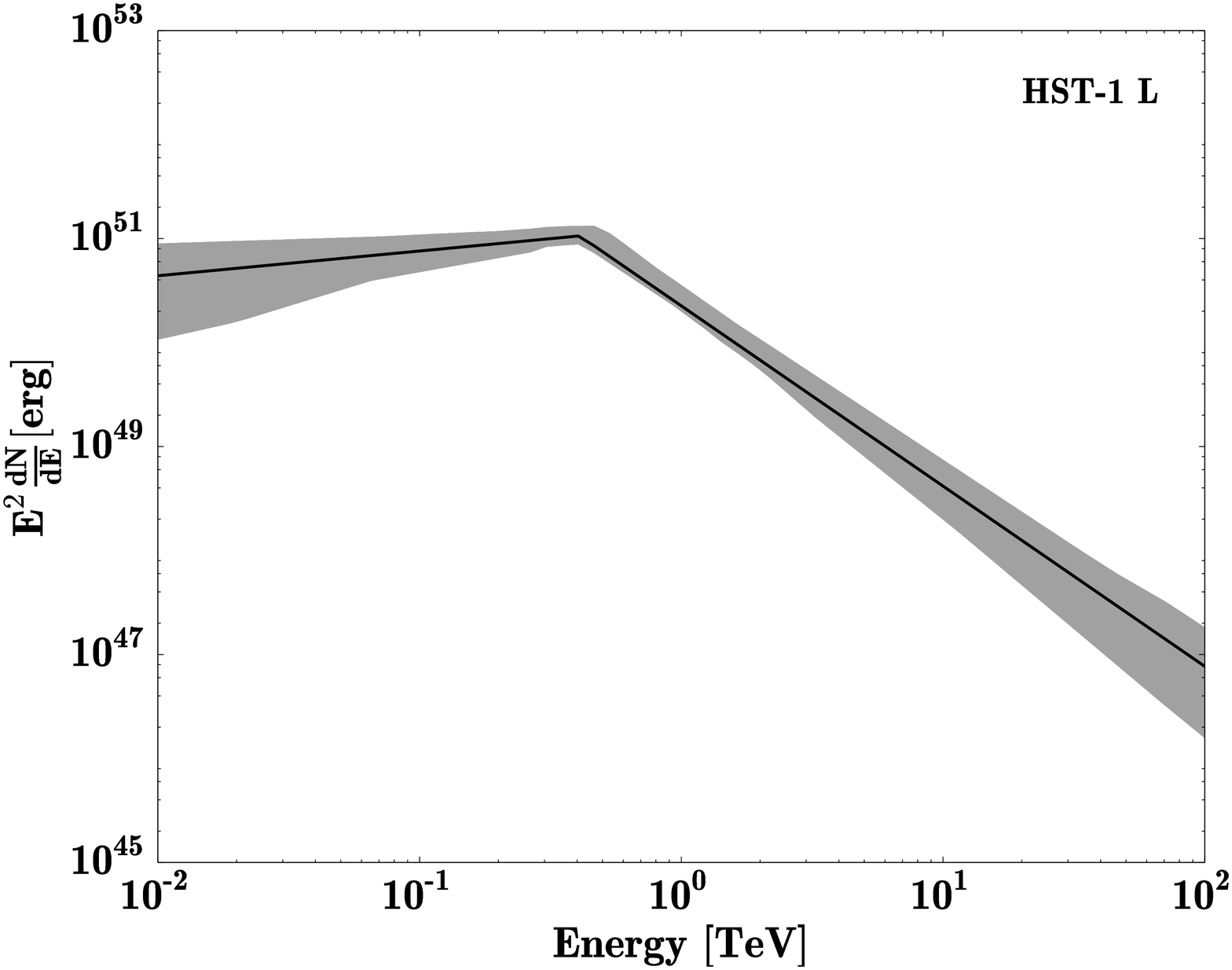}
\includegraphics[width=0.5\textwidth, height=160px]{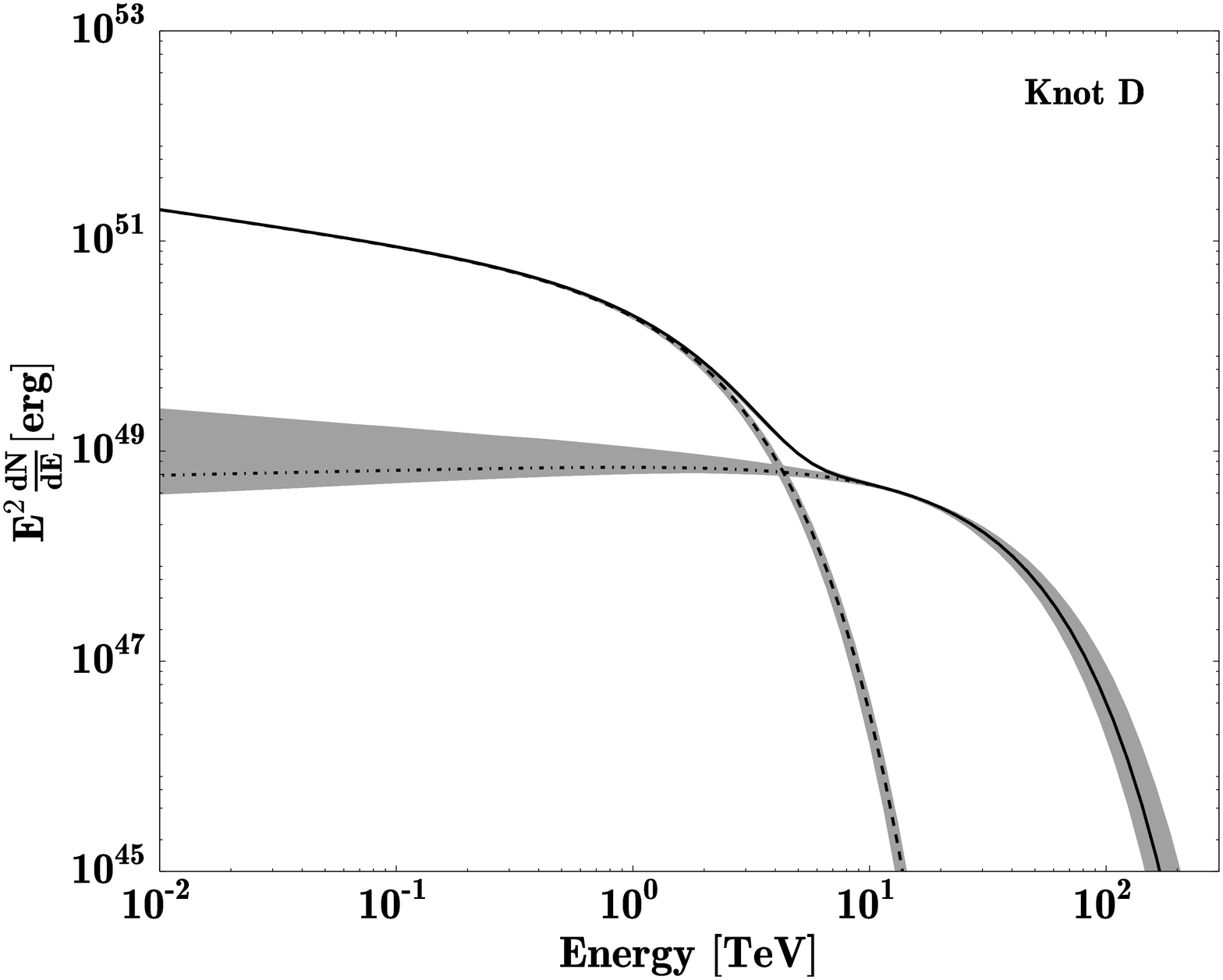}
\includegraphics[width=0.5\textwidth, height=160px]{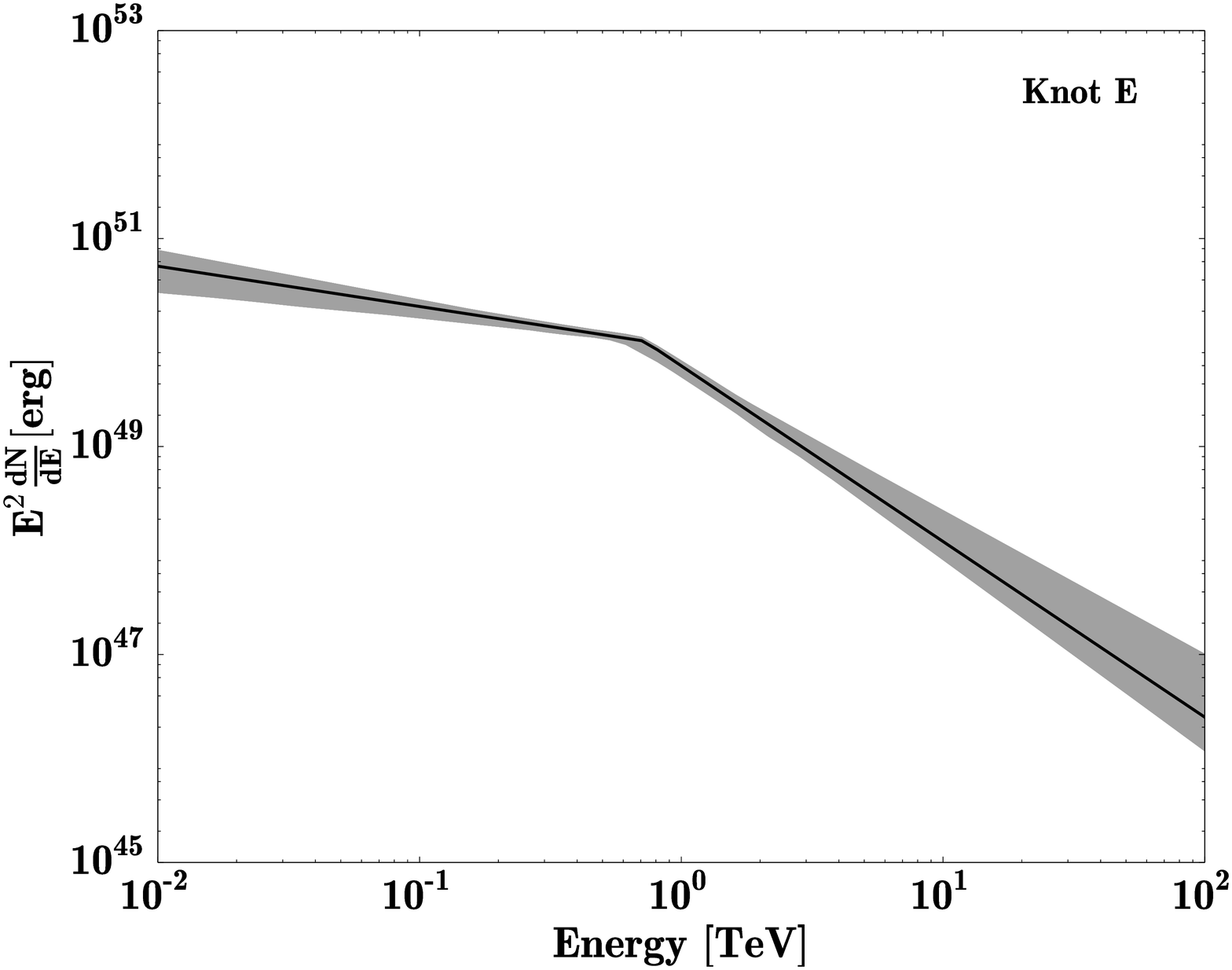}
\includegraphics[width=0.5\textwidth, height=160px]{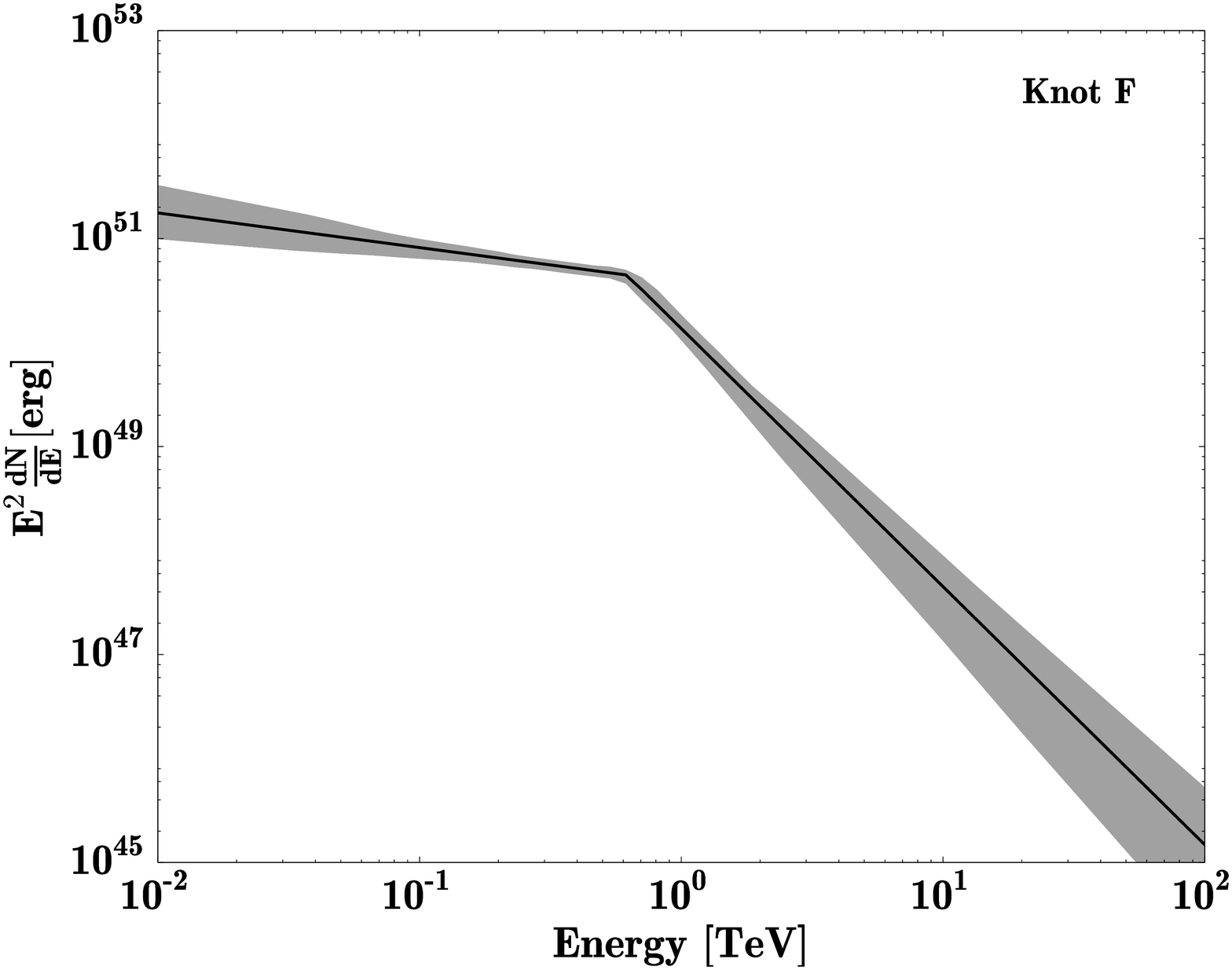}
\includegraphics[width=0.5\textwidth, height=160px]{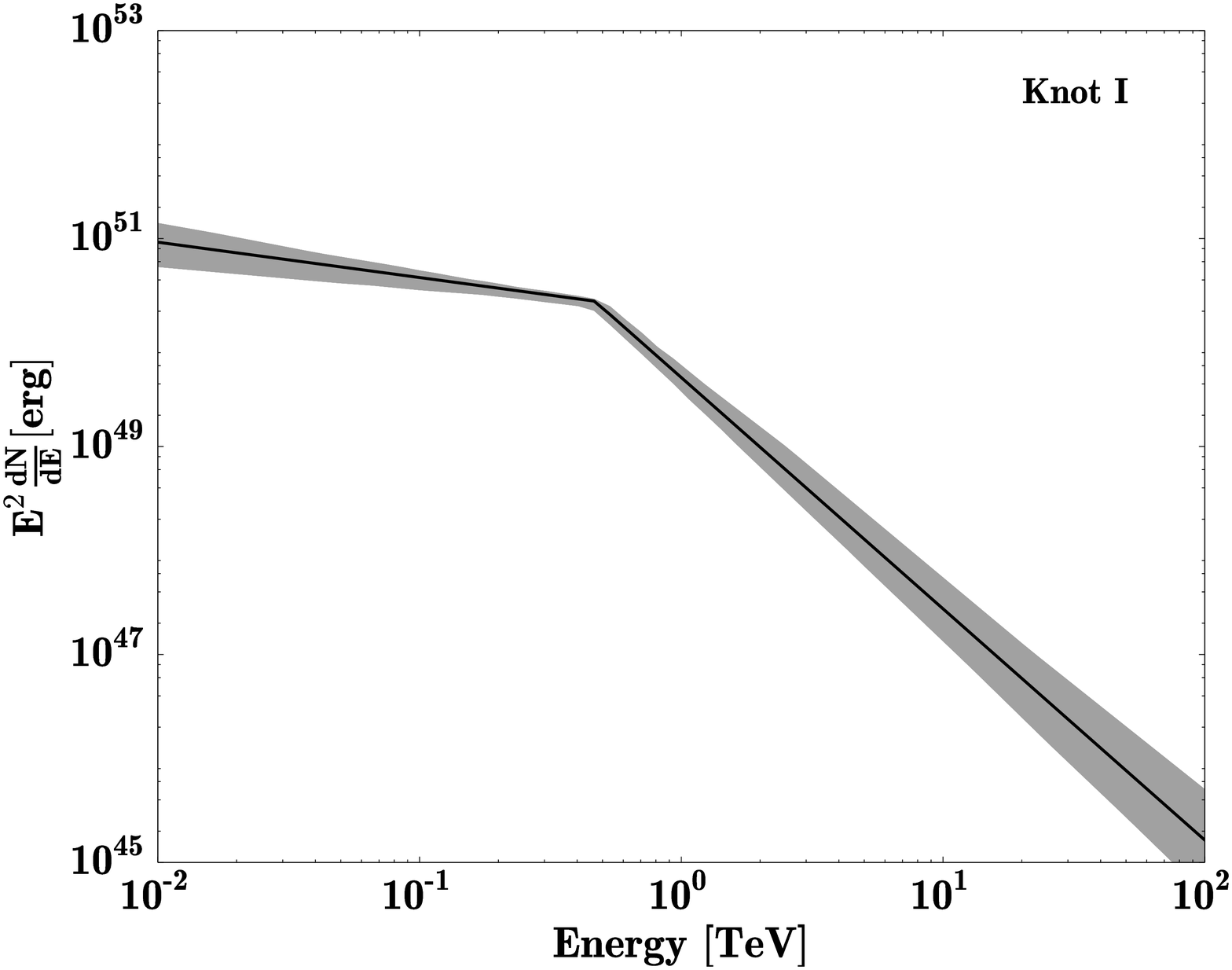}
\includegraphics[width=0.5\textwidth, height=160px]{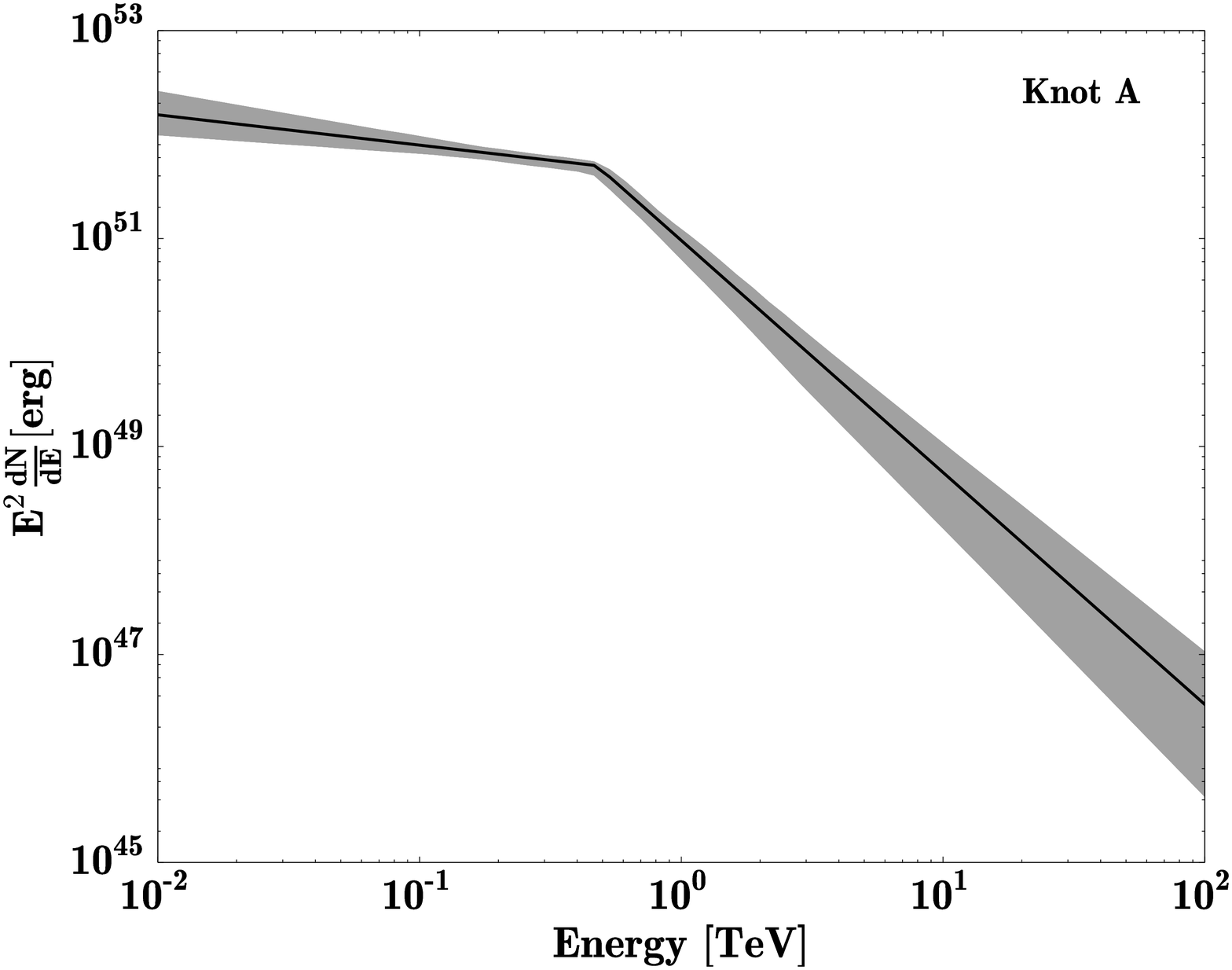}
\caption{Parent electron distributions derived from the fitting in Fig. \ref{fig:sed}. For knots D, B, and C, only the results of the two exponential
cut-off power laws are shown. The grey shadow areas  correspond to a $2\sigma$ confidence level.}
\label{fig:eledis}
\end{figure*}

Our findings provide indications that for  knot regions D, B, and C an additional \xray emitting component is needed to 
achieve a satisfactory SED description. The indications are strongest in the case of knot C where the \xray emission seems 
particularly hard, and less strong in the case of knot D. It is worth noting that radio and optical observations show that the knot 
D region, located approximately between 2 and 5 arcsec, reveals extended structures and can be further divided into 
subregions D-East, D-Middle, and D-West, each with additional subcomponents \citep[e.g.][]{Perlman05,Avachat16}. Similarly, 
knot B, located approximately between 13 and 16 arcsec, can be divided into the subcomponents B1 (broader and brighter) 
and B2, while the knot C region, located approximately between 17 and 19 arcsec, can be divided into the subregions C1 and C2 
\citep{Perlman05,Avachat16}. It seems thus conceivable that the inferred two-component electron distribution can in principle
be accounted for by a superposition of electrons from different subregions. Moreover, the features seen in the optical tend 
to be slightly narrower than those seen in radio, suggesting that the radiating particles, although co-located, may not 
necessarily have to be truly co-spatial. 
We note that we have incorporated only the simplest broken power-law function to fit the SED of the knots (except for B, C, and 
D) and formally cannot rule out the possibility that the SEDs of these knots are produced by more than one electron component 
as well. Noting these caveats, the respective parent electrons distributions are shown in Figure~\ref{fig:eledis}. As can be seen, 
for the considered normalisation of the magnetic field strength the electron distribution tends to show a break at energies around 
one TeV ($\gamma_b \sim 2 \times 10^6$) and needs to extend up to multi-TeV energies ($\gamma \sim 10^8$). These high-energy 
electrons cannot travel more than a few parsecs before exhausting their energies due to synchrotron cooling. A synchrotron origin 
thus requires an efficient and continuous in situ acceleration of electrons along the jet.  

Inverse Compton (IC) up-scattering of cosmic microwave background (CMB) photons ($h\nu_{CMB}$) by these relativistic 
electrons will result in \gray emission extending up to $\sim10~(\gamma/10^8)^2$ TeV. The characteristic energy flux 
levels $\nu F_{\nu}$ at TeV energies are, however, a factor of $\sim u_{CMB}/u_B \sim 10^{-3}$ lower than those seen at 
\xray energies ($\nu F_{\nu} \lesssim 10^{-12}$ erg/cm/s). This TeV emission would thus be below the flux sensitivity limit of 
current \gray instruments.

\clearpage
\clearpage \setlength{\voffset}{0mm}

\begin{figure*}
\includegraphics[angle=0,scale=0.36]{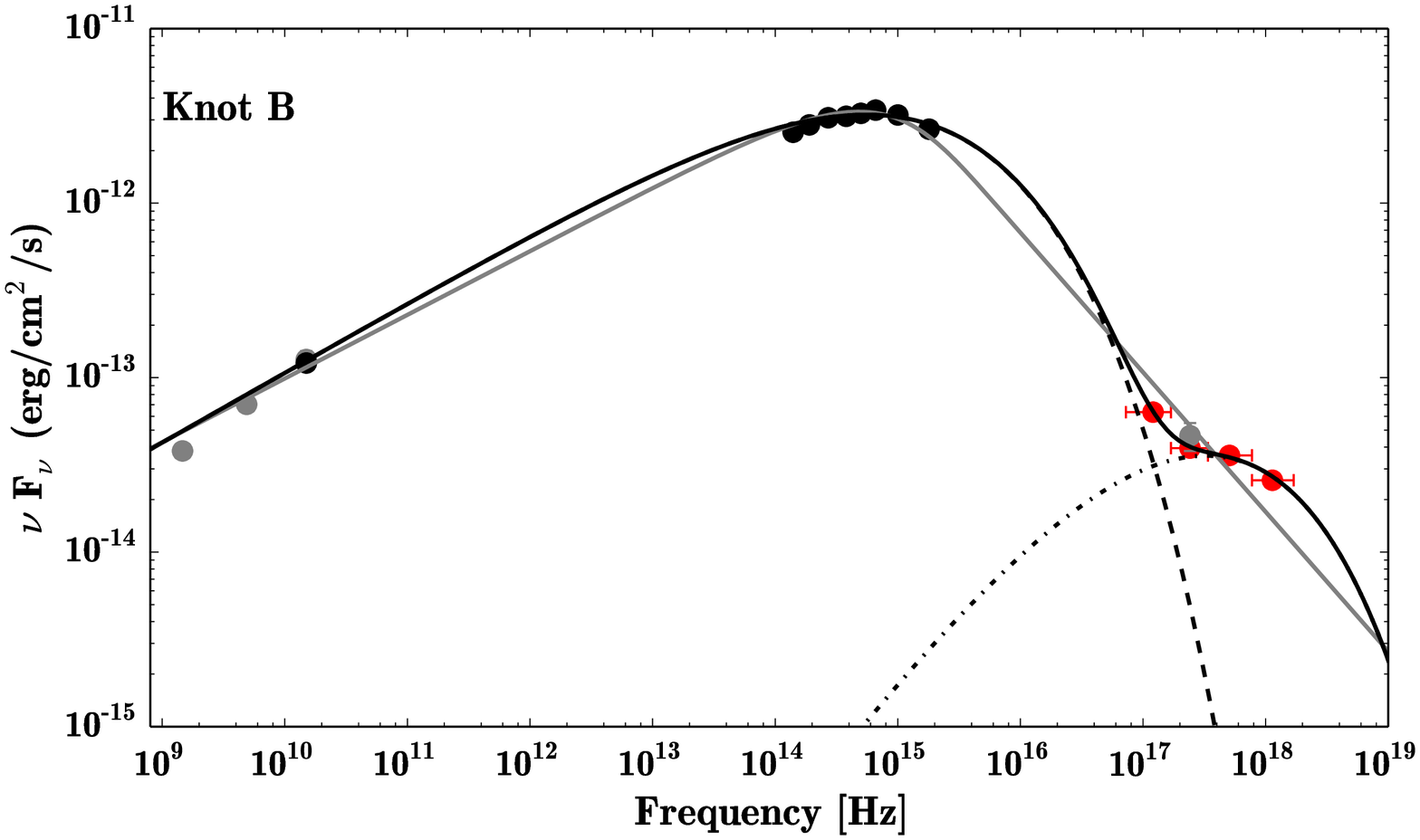}
\includegraphics[angle=0,scale=0.36]{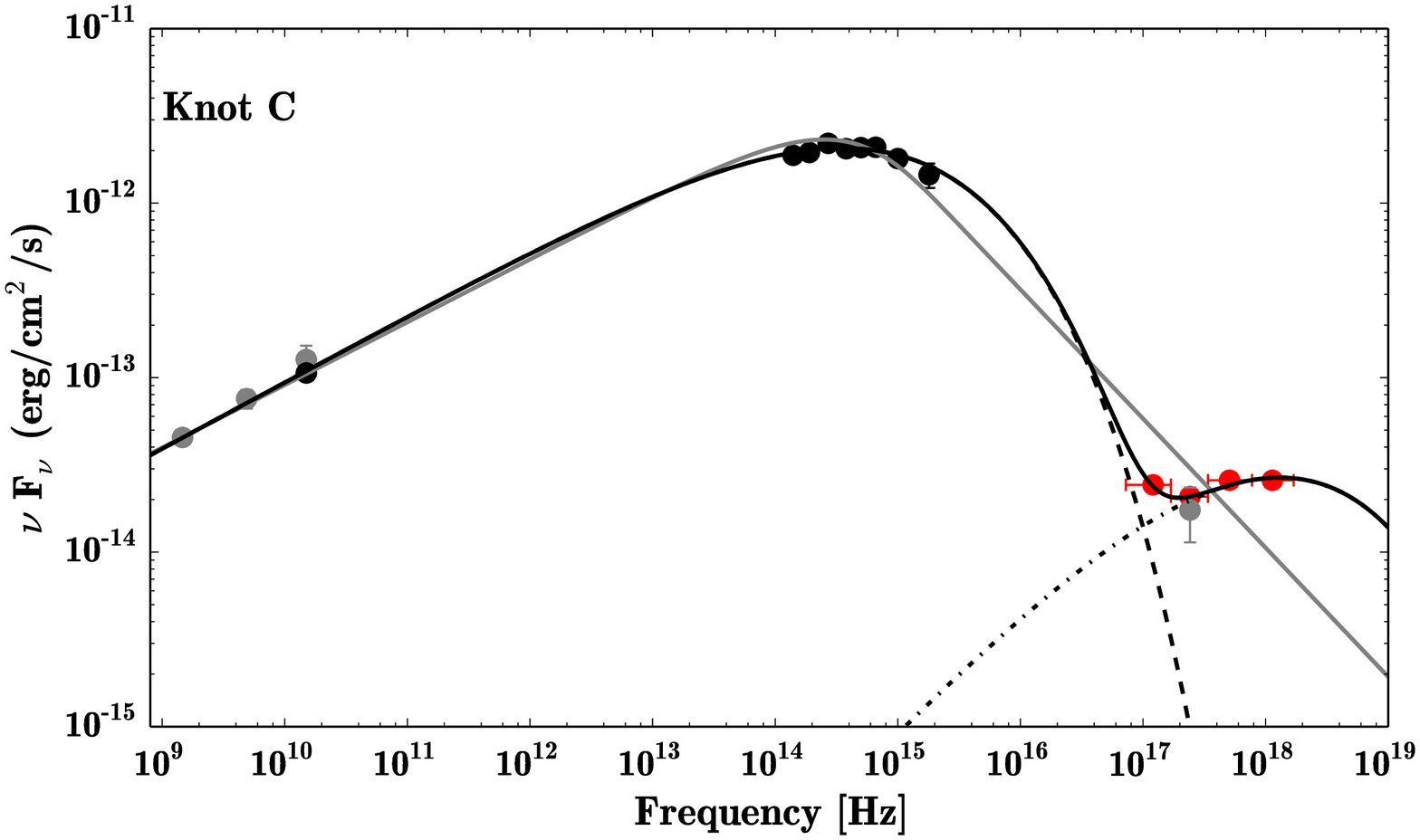}
\hfill\center{Figure ~\ref{fig:sed} ---  continued}
\end{figure*}

\clearpage
\clearpage \setlength{\voffset}{0mm}

\clearpage
\begin{figure*}
\includegraphics[width=0.5\textwidth, height=160px]{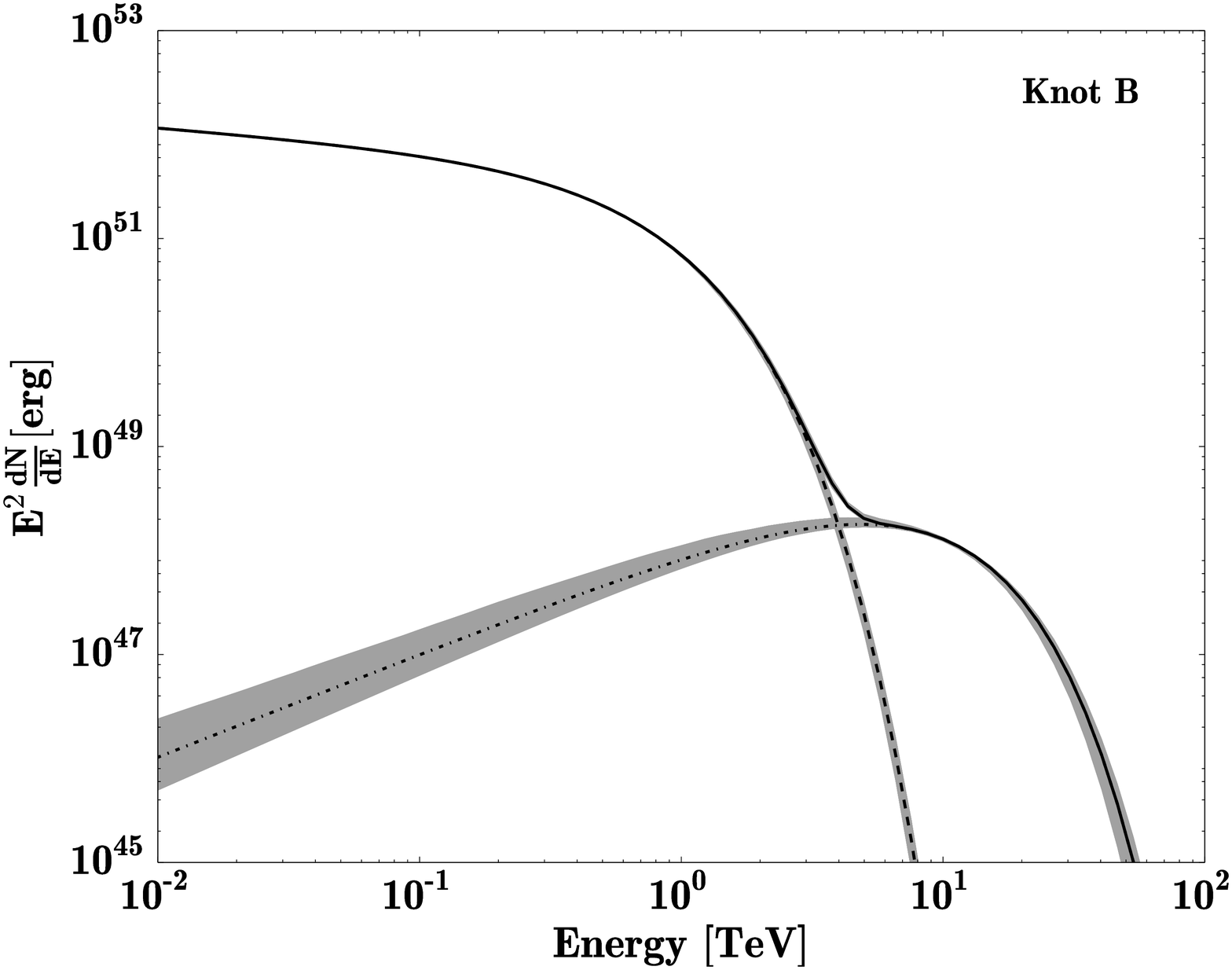}
\includegraphics[width=0.5\textwidth, height=160px]{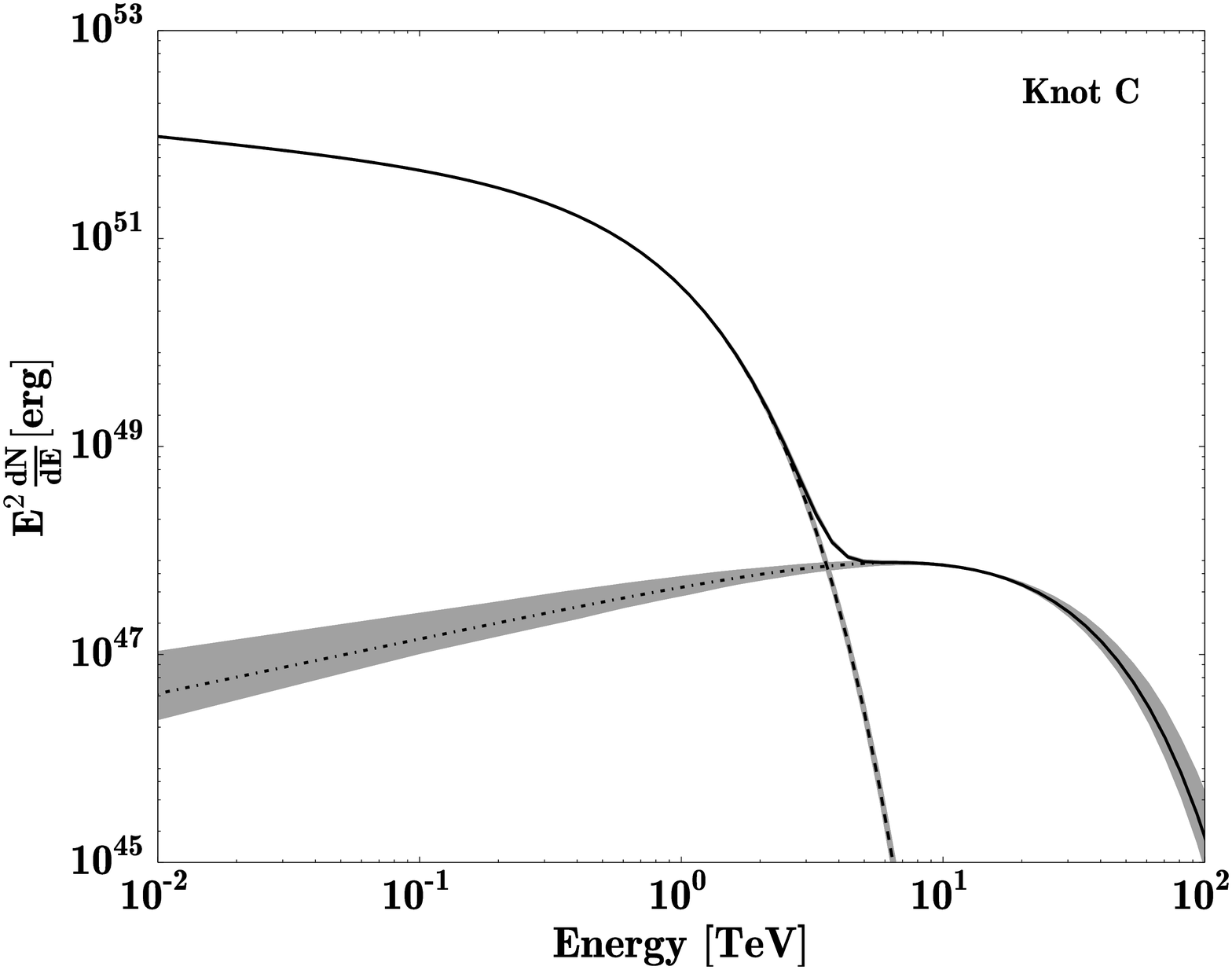}
\hfill\center{Figure ~\ref{fig:eledis} ---  continued}
\end{figure*}

\begin{table*}
\caption{SEDs fit results of the knots of M87. Results of the two-component fit are shown for knots D, B, and C.}
\centering
\begin{tabular}{ccccccc}
\hline\hline
&&Broken power-law model \\ [0.2cm]
\hline
Component & $W_{\rm e} (\rm >1GeV)$ &$\alpha_{1}$ &$E_{\rm break}$ &$\alpha_{2}$&$MLL$\tablefootmark{a}   \\

 &($\times 10^{51}$ erg) & &($\rm GeV$) & \\ [0.1cm]
\hline 
HST-1 $\rm H_{1}$&6.5$^{+1.3}_{-1.9}$&2.08$^{+0.06}_{-0.12}$&1300$\pm$60&3.49$\pm$0.07&-57  \\[0.1cm]
HST-1 $\rm H_{5}$&3.8$\pm$0.5&1.85$^{+0.06}_{-0.09}$&900$^{+60}_{-80}$&3.63$\pm$0.09&-28  \\[0.1cm]
HST-1 L&4.4$^{+0.7}_{-0.5}$&1.77$\pm$0.07&460$^{+20}_{-30}$&3.83$\pm$0.08&-27  \\[0.1cm]
E&3.1$^{+0.4}_{-0.6}$&2.37$^{+0.03}_{-0.06}$&680$^{+40}_{-50}$&3.65$\pm$0.07&-42  \\[0.1cm]
F&10.1$\pm$1.4&2.33$\pm$0.04&630$\pm$30&4.52$\pm$0.10&-79  \\[0.1cm]
I&5.2$\pm$0.8&2.34$^{+0.04}_{-0.03}$&470$\pm$20 &4.23$\pm$0.08&-56 \\[0.1cm]
A&90$^{+14}_{-8}$&2.3$^{+0.04}_{-0.03}$&480$\pm$20&4.25$^{+0.12}_{-0.08}$&-60  \\[0.1cm]
\hline\hline
&&Two exponential cut-off power-laws model\\ [0.2cm]
\hline
Component & $W_{\rm e} (\rm >1GeV)$ &$\alpha_{1}$ &$E_{\rm cutoff1}$&$\alpha_{2}$ &$E_{\rm cutoff2}$&$MLL$\tablefootmark{a} \\
&($\times 10^{51}$ erg) & &($\rm GeV$) & &($\rm GeV$) \\ [0.1cm]
\hline 
D ($\beta$ = 1)&11.3$^{+0.13}_{-0.08}$&2.32$\pm$0.01&1130$\pm$40&2.0$\pm$0.10&19400$^{2000}_{-1700}$&-96 \\[0.1cm]
B ($\beta$ = 1)&62.2$\pm$0.6&2.20$\pm$0.01&521$\pm$9&1.05$\pm$0.05&4900$\pm$200&-11 \\[0.1cm]
C ($\beta$ = 1)&51.2$\pm$0.4&2.24$\pm$0.01&443$\pm$5&1.51$\pm$0.07&13100$\pm$1100&-16 \\[0.1cm]
\hline
D ($\beta$ = 2)&11.1$\pm$1.6&2.36$\pm$0.04&1110$\pm$50&2.14$\pm$0.07&16200$\pm$700&-202 \\[0.1cm]
B ($\beta$ = 2)&58$^{+9}_{-4}$&2.33$\pm$0.03&1150$^{+50}_{-60}$&0.92$\pm$0.04&8000$\pm$400&-49 \\[0.1cm]
C ($\beta$ = 2)&49$\pm$5&2.32$\pm$0.03&640$\pm$30&1.75$^{+0.04}_{-0.05}$&17200$\pm$800&-179 \\[0.1cm]
\hline\hline
&&Log-parabola model \\ [0.2cm]
\hline
Component & $W_{\rm e} (\rm >1GeV)$ &$\alpha$ &$\beta$&$MLL$\tablefootmark{a}   \\

 &($\times 10^{51}$ erg) & & & \\ [0.1cm]
\hline 
HST-1 $\rm H_{1}$&11.5$\pm$1.3&2.35$\pm$0.1&0.12$\pm$0.01&-82  \\[0.1cm]
HST-1 $\rm H_{5}$&7.4$\pm$0.7&2.21$\pm$0.1&0.17$\pm$0.01&-36  \\[0.1cm]
HST-1 L&9.5$^{+0.7}_{-1.3}$&2.36$\pm$0.06&0.23$\pm$0.01&-72  \\[0.1cm]
D&13.5$\pm$1.0&2.67$^{+0.02}_{-0.03}$&0.12$\pm$0.01&-450  \\[0.1cm]
E&4.0$^{+0.4}_{-0.3}$&2.68$^{+0.03}_{-0.02}$&0.12$\pm$0.01&-133  \\[0.1cm]
F&16.4$^{+2.0}_{-1.1}$&2.85$\pm$0.03&0.21$\pm$0.01&-389  \\[0.1cm]
I&7.8$\pm$0.7&2.85$\pm$0.03&0.21$\pm$0.01 &-266 \\[0.1cm]
A&143$\pm$10&2.82$^{+0.05}_{-0.02}$&0.21$\pm$0.01 &-326 \\[0.1cm]
B&106$\pm$9&2.98$\pm$0.04&0.28$\pm$0.01 &-550 \\[0.1cm]
C&84$^{+8}_{-11}$&3.07$^{+0.03}_{-0.06}$&0.28$\pm$0.01 &-2543 \\[0.1cm]
\hline
\end{tabular}
\tablefoot{All errors  correspond to a $1\sigma$ confidence level.
\tablefoottext{a}{Maximum log-likelihood (MML).}
}
\label{table:derived_pra}
\end{table*}

\section{Conclusions}\label{sec:conc}
We have collected \chandra ACIS data for M87 between 2000 and 2016 with a total exposure of about 1.5 Ms to perform a temporal
and spectral analysis of its nucleus and knots. The extracted \xray light curves of the nucleus and HST-1 reveal significant variability. 
We confirm indications for day-scale nuclear \xray variability contemporaneous to the TeV flare in April 2010. HST-1 shows a decline 
in \xray flux since 2007 compatible with its synchrotron origin.

The \xray spectra of the nucleus and jet knots are all formally compatible with a single power law. The resultant \xray photon index 
reveals a trend, with index variations ranging from $\simeq 2.2$ (e.g. in knot D) to $\simeq 2.4-2.6$ (in the outer knots F, A, and B). 
When placed in a multiband context, a more complex situation is seen. Modelling the radio to \xray SEDs with a synchrotron model, 
a single broken power-law electron distribution with a break at around $E_b \sim 1$ TeV (assuming $B\sim 300 \mu$G) seemingly allows 
a satisfactorily SED description for most knots. However, for  knots B, C, and D an additional high-energy component is needed 
to account for the broad-band SEDs. This may be partly due to a blending of different (non-resolved) components and/or the 
occurrence of additional acceleration and emission processes. 

The most  favourable interpretation for the origin of the \xray emission is synchrotron radiation of relativistic electrons. This requires that
electrons should be able to reach and sustain energies of $\gamma \sim 10^8~(300\mu \mathrm{G}/B)^{1/2}$ in the presence of losses,
i.e. an in situ acceleration of electrons in the jet and its knots. The necessary acceleration efficiency $\eta$, defined by $t_{\rm acc} =
\gamma m_e c/(\eta eB)$ \citep{Aharonian02b}, would be of the order of $\eta \gtrsim 3\times 10^{-4}~(B/300\mu\mathrm{G})$, and 
likely to be in the reach of stochastic acceleration scenarios \citep[e.g.][]{Rieger07}. Localised (first-order Fermi) shock acceleration
alone, however, would not be sufficient given  that here is little evidence for  the inter-knot regions to have significantly 
steeper spectra than the adjacent knots.

The homogeneous broken power-law model exhibits a change in indices exceeding that induced by simple cooling effects. 
Assuming a continuous power-law injection, \citet{Perlman05} have suggested that an energy-dependent filling factor 
$f_{acc}(\gamma)\propto \gamma^{-\xi}$ of the acceleration regions (with $\xi \sim 0.3$, i.e. occupying a smaller fraction at 
higher energies), might account for these breaks.
Alternative explanations could be a spatial varied propagation of the relativistic electrons or a different particle injection spectrum 
in a different energy band.  However, to some extent all of these assumptions are ad hoc, and detailed modelling as well as an 
advanced morphological analysis in different energy bands would be required to further qualify them. 

There are clear indications that in the case of knots B, C, and D an additional electron contribution is needed 
to account for the \xray emission in a multiband context. We formally cannot rule out the possibility that the SEDs of other knots 
are also produced by a two-component (or more) electron distribution, though the smoothness of the fits might seem to argue for 
the opposite, i.e. there is no evidence that the \xray emission in these knots  consists of separate spectral components.

The additional electron component indicated in the case of knots B, C, and D is consistent with a bump or a spike-like feature at 
high energy.  This appears compatible with a scenario where electron acceleration occurs in a shear layer and radiative loss is 
effective \citep{Ostrowski00}. The short cooling time of the relativistic electrons generally requires that in addition to a possible
localised shock-type scenario there must be some distributed in situ acceleration occurring along the jet. Stochastic or shear 
acceleration could present a natural explanation for this \citep[e.g.][]{Liu17}. There is circumstantial evidence for a decreased 
thermal gas emissivity along the jet that appears consistent with the dynamical effects of a cocoon dominated by cosmic rays (CR) 
accelerated at the shearing jet side boundary \citep{Dainotti12}. 
The hint of a CR cocoon and the spectral shape of the relativistic electrons in these knots could indicate that efficient shear 
acceleration indeed is taking place along the jet in M87, which seems to link well into the broader evidence for a stratified or
spine-sheath flow \cite[e.g.][]{Perlman99}.

\section*{Acknowledgements}
We thank the referee Eric Perlman for the very useful comments and suggestions that helped to improve the paper.
X.N. Sun thanks Jin Zhang for helpful discussions, and acknowledges the financial support of the China Scholar Council.
F.M.R. kindly acknowledges support by a DFG Heisenberg Fellowship (RI 1187/4-1). 
This research has made use of data obtained from the \chandra Data Archive and the \chandra Source Catalog, and 
software provided by the \chandra \xray Center (CXC) in the application packages CIAO, ChIPS, and Sherpa. 

\bibliographystyle{aa}
\bibliography{ms_v6}

\end{document}